\documentclass[12 pt]{article}
\usepackage{bbm}
\usepackage[longnamesfirst]{natbib}
\usepackage{amsmath}
\usepackage{amssymb}
\usepackage{amsthm}
\usepackage{setspace}
\usepackage[colorlinks]{hyperref}
\usepackage{graphicx}
\usepackage{caption}
\usepackage{enumitem}
\setlist{noitemsep,parsep=6pt,partopsep=0pt,topsep=0pt}
\usepackage[margin=1in]{geometry}
\usepackage{verbatim}
\usepackage{sectsty}
\usepackage[svgnames]{xcolor}
\usepackage{subfiles}
\usepackage{xargs}
\usepackage[titletoc,title]{appendix}
\usepackage{titlesec}
\usepackage[bottom]{footmisc}
\usepackage[multiple]{fnpct}
\usepackage[colorinlistoftodos,textsize=tiny]{todonotes}
\usepackage[capitalize,nameinlink]{cleveref}
\usepackage[normalem]{ulem}
\usepackage[font={small,it},justification=justified]{caption}
\usepackage{tablefootnote}
\usepackage{threeparttable}
\usepackage{afterpage}
\usepackage{xparse}
\hypersetup{
    pdftitle={How Wasteful is Signaling?},
    pdfauthor={Alex Frankel and Navin Kartik},
    citecolor=DarkBlue,
    bookmarksnumbered=true,
    urlcolor=Indigo,
    linkcolor=DarkBlue,
}
\usepackage{subcaption,needspace}
\usepackage{xspace}
\usepackage{booktabs}

\renewcommand{\epsilon}{\varepsilon}

\newcommand{\appendixref}[1]{\hyperref[#1]{Appendix \ref{#1}}}
\newcommand{\onlineappendixref}[1]{\hyperref[#1]{Supplementary Appendix \ref{#1}}}

\newtheorem{theorem}{Theorem}
\newtheorem{lemma}{Lemma}

\newtheorem{proposition}{Proposition}

\newtheorem{assumption}{Assumption}

\theoremstyle{remark}
\newtheorem{remark}{Remark}

\theoremstyle{definition}
\newtheorem{example}{Example}
\makeatletter
\AtBeginEnvironment{example}{%
  \pushQED{\qed}%
}
\AtEndEnvironment{example}{%
  \popQED
}
\makeatother

\newtheorem{definition}{Definition}
\NewDocumentCommand{\citepos}{sm}{%
  \IfBooleanTF{#1}
    {\citeauthor*{#2}'s~\citeyearpar{#2}}  
    {\citeauthor{#2}'s~\citeyearpar{#2}}   
}
\newcommand{\R}{\mathbb{R}}
\newcommand{\Rpos}{\R_{\geq0}}
\newcommand{\Rstrpos}{\R_{>0}}

\newcommand{\E}{\mathbb{E}}
\newcommand{\Prob}{\mathbb{P}}

\def\typemin{0}
\def\typemax{\overline \theta}
\newcommand{\elaB}{\tilde\beta}   
\newcommand{\elaS}{\tilde\sigma}  
\newcommand{\elaR}{\tilde\rho}    

\usetikzlibrary{patterns, decorations.pathreplacing, positioning, arrows.meta,calc,decorations.pathreplacing,calligraphy,shapes.geometric,decorations.markings}

\let \savenumberline \numberline
\def \numberline#1{\savenumberline{#1.}}

\makeatletter
  \renewcommand\@seccntformat[1]{\csname the#1\endcsname.{\hskip.7em\relax}} 
\makeatother
\makeatletter
\renewenvironment{proof}[1][\proofname] {\par\pushQED{\qed}\normalfont\topsep6\p@\@plus6\p@\relax\trivlist\item[\hskip\labelsep\bfseries#1\@addpunct{.}]\ignorespaces}{\popQED\endtrivlist\@endpefalse}
\makeatother

\newcommand{\mailto}[1]{\href{mailto:#1}{\texttt{#1}}} 

\newcommand{\suppapp}[1]{\hyperref[#1]{Supplementary Appendix}}

\let\oldfootnote\footnote
\renewcommand\footnote[1]{\oldfootnote{\hspace{.5mm}#1}}

\titlespacing\section{0pt}{10pt plus 2pt minus 2pt}{4pt plus 2pt minus 2pt} 
\titlespacing\subsection{0pt}{6pt plus 2pt minus 2pt}{2pt plus 2pt minus 2pt} 
\titlespacing\subsubsection{0pt}{6pt plus 2pt minus 2pt}{0pt plus 2pt minus 2pt} 
\titlespacing{\paragraph}{0pt}{0.5\baselineskip}{1em}

\usepackage{xcolor}

\sectionfont{\color{DarkRed}}
\subsectionfont{\color{DarkRed}}
\subsubsectionfont{\color{DarkRed}}

\begin{document}

\title{How Wasteful is Signaling?\thanks{We thank Nageeb Ali, Emir Kamenica, Hongcheng Li, Elliot Lipnowski, George Mailath, Benny Moldovanu, Georg N{\"o}ldeke, Andrea Prat, \href{https://www.refine.ink}{Refine.ink}, Larry Samuelson, and Joel Sobel for helpful comments.}}

\author{Alex Frankel\thanks{University of Chicago, Booth School of Business; Email: \mailto{afrankel@chicagobooth.edu}.} \and Navin Kartik\thanks{Yale University, Department of Economics; Email: \mailto{nkartik@gmail.com}.}}

\maketitle
\thispagestyle{empty}

\begin{abstract}
\noindent
Signaling is wasteful. But how wasteful? We study the fraction of surplus dissipated in a separating equilibrium. For isoelastic environments, this waste ratio has a simple formula: $\beta/(\beta+\sigma)$, where $\beta$ is the benefit elasticity (reward to higher perception) and $\sigma$ is the elasticity of higher types' relative cost advantage. The ratio is constant across types and is independent of other parameters, including convexity of cost in the signal. We show that the directional effects of $\beta$ and $\sigma$ on waste extend to non-isoelastic environments. In an application to signaling tournaments, more competitors or fewer prizes increase waste, with full dissipation in large tournaments.
\end{abstract}

\newpage
\setcounter{page}{1}

\section{Introduction}
\label{sec:intro}

Signaling is wasteful. In the canonical \citet{Spence73} model and its innumerable applications and descendants, agents take costly actions to distinguish themselves from lower types. The resulting separating equilibrium reveals information but necessarily dissipates surplus---a fundamental source of inefficiency under asymmetric information.\footnote{\label{fn:productive}Of course, signaling activities can also generate benefits: education builds human capital; and prosocial behavior brings positive externalities, which signaling can amplify \citep{BT06}. Our paper  focuses on the wasteful component of signaling.} 

But how wasteful is signaling, and what does the waste depend on? Despite more than 50 years of research, these basic questions have received limited attention, and the literature does not offer simple answers.

A natural intuition suggests that the magnitude of waste should depend on the \emph{difficulty} of signaling. If signaling costs are highly convex in the action (a ``hard'' test), agents encounter high marginal costs quickly, which ought to limit total expenditure. This reasoning suggests that policies that make signaling more difficult---via exam difficulty, advertising costs, or certification requirements---could reduce waste. However, while such policies reduce the level of signaling, they also increase the cost of lower signals.

Reducing signaling \emph{stakes} instead---scaling down the benefits of being thought of as a higher type---also dampens signaling, and does indeed lower signaling costs. But reducing stakes also lowers signaling benefits. 
It is not obvious what the overall effect of difficulty or stakes is on the \emph{waste ratio}, i.e., the proportion of private surplus burned through signaling.

Our paper studies the classic continuum-type signaling model used in economics, presented formally in \autoref{sec:model}, and focuses on the essentially unique separating equilibrium. Our main result, \autoref{thm:constant} in \autoref{sec:results}, has two parts. First, under a standard multiplicative cost structure, the waste ratio is invariant to both difficulty and stakes. Importantly, difficulty captures not just the scale of costs, but also the shape (convexity).

Second, consider a canonical isoelastic class of costs and benefits: the cost for type $\theta$ of taking signaling action $a$ is given by $C(a,\theta)=D(a) \theta^{-\sigma}$, while the benefit of being thought of as type $\hat \theta$ is $V(\hat \theta) = s \hat \theta^\beta$. Here, $\beta>0$ is the elasticity of benefits (how steeply rewards rise with perceived type),  $\sigma>0$ is the elasticity of cost ``strain'' (how quickly higher types' comparative advantage grows), and $D(\cdot)$ and $s>0$ are the difficulty and stakes respectively. Under such isoelasticity, we find that the waste ratio for any type is the constant 
\[
W = \frac{\beta}{\beta + \sigma}.
\]
Waste thus depends only on $\beta/\sigma$, increasing from zero to one in that fraction. This constant waste ratio avoids issues of aggregation across types and delivers a simple answer to our motivating question. 

We then show in \autoref{thm:char} that (under multiplicative costs) waste is constant across types {if and only if} the costs and benefits satisfy a constant relative elasticity condition. Hence, up to the labeling of types, the isoelastic class is the unique setting for such uniform dissipation. This characterization provides a theoretical foundation for the isoelastic specification. 

\autoref{prop:comp_stat} establishes comparative statics beyond the isoelastic specification: when benefit and cost strain elasticities are type dependent, a pointwise increase in the benefit elasticity and pointwise decrease in the strain elasticity imply a higher waste at every type. This generalizes the directional effects of $\beta$ and $\sigma$ seen in the isoelastic waste formula.

In \autoref{sec:welfare}, we relate the waste ratio, which accounts only for agents’ private benefits from signaling, to the social value of information. Specifically, we consider a labor market in which workers signal to competing firms that make productive investments complementary to the type of worker they hire. We compare the separating equilibrium to a pooling equilibrium, which avoids waste but has less efficient investments. The efficiency benefits of separation outweigh the signaling costs when the type distribution is sufficiently spread out or when the signaling technology has high strain elasticity.

In \autoref{sec:tournament}, we apply our results to signaling in tournaments. When $N$ candidates compete for $m$ prizes, the waste increases in $N$ and decreases in $m$. In other words, greater competition for prizes yields higher rent dissipation. With symmetric agents and any fixed number of prizes, a growing field eventually dissipates all surplus. One special case is with uniformly distributed types, unit strain elasticity, and a winner-take-all market with $m=1$. In that case, the waste ratio takes the simple form of $(N-1) / N$, which is precisely the dissipation rate in classic \citet{Tullock80} contests.

The conclusion, \autoref{sec:conclusion}, discusses implications, interpretations, and limitations.

\paragraph{Related Literature.} The costly signaling literature in economics, surveyed by \citet{Riley01} and \citet{Sobel09}, identifies conditions for separating equilibria and highlights that information revelation entails surplus dissipation.  However, we are aware of little work that systematically analyzes this waste. A notable exception is \citet*{HoppeMoldovanuSela09}, who show that in a continuum model of two-sided signaling before assortative matching, exactly half of total output is dissipated. Viewed agent by agent, their model in fact maps into ours, with the one-half quantity emerging from our waste formula under their parameters; \autoref{fn:hms-translation} elaborates. Another exception is \citet{BBC23}, who show that dissipation can vanish when agents have heterogeneous bliss points and choose many actions (or, equivalently, costs are scaled up). Our paper instead quantifies waste in the canonical signaling setting with homogeneous bliss points.

In the biological signaling literature, \citet{NoldekeSamuelson99} show that offspring's equilibrium cost is proportional to parents' fitness loss, with a constant depending only on genetic relatedness. Their analysis does not yield a constant waste ratio (cost relative to sender's benefit, which need not track parental loss), and their assumption of a linear cost precludes questions about signaling difficulty. But our paper shares with them a common theme that given some structure, dissipation can admit a simple formula based on primitive parameters, with certain invariance properties. By contrast, \citet{BSL02} point out that without structure, little can be said about the equilibrium level of signaling costs.

We discuss some other literature connections later in the paper.

\section{Model}
\label{sec:model}

An agent has type $\theta\in \Theta:=[\typemin,\typemax\rangle$, where $0<\typemax \leq \infty$.\footnote{We use the notation $[\typemin,x\rangle$ to mean $[\typemin,x]$ if $x<\infty$ and $[0,\infty)$ if $x=\infty$.}  The type is drawn from some distribution with full support on $\Theta$. After privately learning her type, the agent chooses a publicly observable signal or action $a \in \Rpos$. An observer sees the action and forms a belief $\hat \theta \in \Theta$ about the agent's type.\footnote{As we will focus on separating equilibria, we only need to consider degenerate beliefs on a single type.} The agent's payoff is $
V(\hat\theta) - C(a, \theta)$. We maintain throughout the following assumption (primes and subscripts on functions denote derivatives in the usual manner).

\begin{assumption}The benefit function $V: \Theta \to \Rpos$ and cost function $C: \Rpos \times \Theta \to \Rpos \cup \{\infty\}$ satisfy:
\label{ass:basics}
\begin{enumerate}
\item \label{ass:V} $V$ is continuous, with $V(\typemin) = 0$; on the interior $(\typemin,\typemax)$, it is continuously differentiable with $V'(\theta) > 0$.
\item \label{ass:C} On $\Rpos \times (\typemin,\typemax\rangle$, $C$ is finite and continuous with $C(0,\theta)=0$; on $\Rstrpos \times (\typemin,\typemax\rangle$, $C$ is differentiable in $a$ with $C_a>0$, and $C_a$ is continuously differentiable with $C_{a\theta} < 0$;  and for each $\theta>\typemin$, $\lim_{a\to\infty} C(a,\theta)=\infty$. The lowest type has cost $C(a,\typemin)=\lim_{\theta \rightarrow \typemin} C(a, \theta)$ for all $a$.
\end{enumerate}
\end{assumption}

Part \ref{ass:V} of \autoref{ass:basics} says that agents prefer to be perceived as higher types, with the benefit from the lowest perception normalized to zero. Part \ref{ass:C} says that higher actions are costlier, and higher types have lower marginal costs. While it may be natural for costs to be convex in the action, we don't need to assume that. Part \ref{ass:C} also normalizes $C(0,\theta) = 0$ for all $\theta$, so that the payoff from taking the lowest action and receiving the lowest perception is zero. The technical conditions in the two parts are largely standard; note that we allow for type $\typemin$ to have infinite costs for actions $a>0$ to encompass canonical isoelastic costs, detailed in \autoref{sec:results}.

\paragraph{Equilibrium.} We study (fully) \emph{separating equilibria}. The equilibrium definition is standard and relegated to \appendixref{sec:separating}, where \autoref{prop:separating-differentiable} shows that any separating equilibrium can be described by a continuous, strictly increasing agent (pure) strategy $A: \Theta \to \Rpos$ that is differentiable at interior types and satisfies $A(\typemin) = 0$.  Incentive compatibility requires that each type $\theta$ optimally chooses $A(\theta)$ given that the observer correctly inverts the strategy on the equilibrium path, i.e., when beliefs satisfy $\hat\theta(a) = A^{-1}(a)$ in the range of $A$. Off-path beliefs can simply be set to $\hat \theta(\cdot)=\typemin$. 

Thus, in a separating equilibrium $A$, any type $\theta$ solves
\[
\max_a \; [V(\hat\theta(a)) - C(a, \theta)],
\]
where $\hat \theta(\cdot)=A^{-1}(\cdot)$. 
For $\theta\in(\typemin,\typemax)$, the first-order condition evaluated at the optimal action $A(\theta)$ is
\begin{equation}\label{e:foc}
C_a(A(\theta), \theta) = V'(\theta) \cdot \hat\theta'(A(\theta)) = \frac{V'(\theta)}{A'(\theta)},
\end{equation}
where the first equality uses $\hat \theta(A(\theta))=\theta$ and the second uses $\hat\theta'(A(\theta)) = 1/A'(\theta)$.

\autoref{e:foc} has a simple interpretation. Its left-hand side is the marginal cost of increasing the action; the right-hand side is the marginal benefit of inducing a higher belief scaled by the marginal increase in action required for that higher belief. Together with $A(\typemin) = 0$, \autoref{e:foc} defines a boundary-value differential equation in $A$. There is a unique solution by standard arguments.\footnote{More precisely, standard existence and uniqueness results for ordinary differential equations can be applied on $(\typemin,\typemax)$ and extended to the boundary by continuity; see the arguments in, for example, \citet{Mailath87} or \citet{Kartik09}.} That solution, which we continue to refer to as just $A$ subsequently, characterizes the unique separating equilibrium (uniqueness is up to the specification of off-path beliefs); sufficiency is verified by \autoref{prop:sufficiency} in \autoref{sec:separating}.

\paragraph{The Waste Ratio.} To measure signaling inefficiency we define three quantities. The {opt-out} payoff $U^{O}(\theta):=V(\typemin)-C(0,\theta)=0$ is what a type would get if it chose the least-cost action and was perceived as the lowest type. The {complete-information} payoff $U^{CI}(\theta):=V(\theta)-C(0,\theta)=V(\theta)$ is what a type would get if it revealed itself costlessly. Lastly, $U(\theta):=V(\theta) - C(A(\theta), \theta)$ is a type's separating {equilibrium payoff}.

\begin{definition}
\label{def:waste}
The \emph{waste ratio} for type $\theta>0$ is the fraction of its payoff from costless separation that is dissipated through costly signaling:%
\begin{equation}
\label{e:waste}    
W(\theta) := \frac{U^{CI}(\theta) - U(\theta)}{U^{CI}(\theta) - U^O(\theta)} = \frac{C(A(\theta), \theta)}{V(\theta)}.
\end{equation}
\end{definition}

We refer to the denominator $V(\theta)$ as \emph{surplus}: it is the payoff that type $\theta$ would hypothetically get by verifying her type at zero cost. The numerator $C(A(\theta), \theta)$ is the deadweight loss from signaling. The ratio $W(\theta)$ thus measures the effective ``tax'' that the separating equilibrium imposes on the agent to secure her surplus.

Note that our definition of waste compares the agent's cost of information revelation to a frictionless benchmark in which information is revealed at no private cost. This benchmark is, of course, unachievable. Relatedly, we are not defining waste relative to a pooling equilibrium or any other equilibrium. Waste is also only defined in terms of the agent's private surplus, not necessarily social surplus from information. 
We discuss social surplus and pooling equilibria in \autoref{sec:welfare}.

Our goal is to understand how the waste ratio \eqref{e:waste} depends on the parameters of the signaling environment.\footnote{The waste ratio can be viewed as analogous to the ``Price of Anarchy'' in algorithmic game theory \citep{KP99, Roughgarden05}. That literature generally studies worst-case bounds across multiple equilibria; we are interested in the exact value in the separating equilibrium. Furthermore, we define waste pointwise across types, whereas Bayesian Price of Anarchy typically uses ex-ante expected payoffs \citep{RST17}. A consequence of our results is that the latter distinction is rendered moot in \hyperref[def:isoelastic]{isoelastic environments}.}

\section{Signaling's Waste}
\label{sec:results}

We hereafter focus on multiplicatively separable costs---or just multiplicative costs, for short---that are commonplace in signaling models. Formally, we assume that
\begin{equation}\label{e:multsep}
C(a, \theta) = D(a) \cdot S(\theta),
\end{equation}
where $D: \Rpos \to \Rpos$ and $S:\Theta \to \Rstrpos \cup \{\infty\}$. Here $D(a)$ represents the \emph{difficulty} of action $a$ (relative to other actions) and $S(\theta)$ represents the \emph{strain} experienced by type $\theta$ (relative to other types). \autoref{ass:basics} part \ref{ass:C} implies (i) $D(0) = 0$, $D'(a) > 0$ for $a > 0$, and $\lim_{a\to\infty}D(a)=\infty$; and (ii) for $\theta>0$, we have $S(\theta)$ finite and $S'(\theta)<0$, while $S(0)=\lim_{\theta\rightarrow 0}S(\theta)$. Note that $S(0) = \infty$ corresponds to type $0$ facing prohibitive signaling costs for any $a>0$.\footnote{But $C(0,0)=D(0)S(0)=0$, using the convention $0\times \infty =0$.}

It is also useful to write, without loss, $$V(\theta)= s\cdot B(\theta),$$ where $s>0$ represents the agent's \emph{stakes} in signaling and $B:\Theta\to \Rpos$.

We define, for interior $\theta$, the \emph{benefit elasticity}
\[
\elaB(\theta) := \frac{d\ln V(\theta)}{d\ln\theta} = \frac{d\ln B(\theta)}{d\ln\theta} = \theta\frac{B'(\theta)}{B(\theta)}>0,
\]
and, for multiplicative costs, the \emph{strain elasticity}
\[
\elaS(\theta) := -\frac{\partial\ln C(a,\theta)}{\partial\ln\theta} = -\frac{d\ln S(\theta)}{d\ln\theta} = -\theta\frac{S'(\theta)}{S(\theta)}>0.
\]
This pair of elasticity functions characterizes an environment under multiplicative costs: given $\elaB$ and $\elaS$, one can recover $B$ and $S$ up to normalization constants.

\subsection{The Constant of Dissipation}
\label{subsec:constant}

A leading parametric specification is that of isoelastic costs and benefits:

\begin{definition}
\label{def:isoelastic}
An \emph{isoelastic environment} is defined by
\[
B(\theta) = \theta^\beta \quad \text{and} \quad S(\theta) = \theta^{-\sigma},
\]
for some constant {benefit elasticity} $\beta > 0$ and constant {strain elasticity} $\sigma > 0$.
\end{definition}

Note that the definition stipulates isoelasticity in $B$ and $S$, but not in difficulty $D$. Our first result says that multiplicative costs ensure the waste ratio is independent of stakes and difficulty, and isoelasticity further implies a constant waste ratio across types.

\begin{theorem}
\label{thm:constant}
Under multiplicative costs:
\begin{enumerate}
\item \label{invariance} The waste ratio $W(\theta)$ is invariant to stakes ($s$) and difficulty ($D(\cdot)$).
\item \label{constant} In an isoelastic environment, the waste ratio is constant: for any $\theta>0$, it is
\begin{equation}
\label{e:constant}
W(\theta) = \frac{\beta}{\beta + \sigma}.    
\end{equation}
\end{enumerate}
\end{theorem}

The irrelevance of the difficulty $D$ in the first part of \autoref{thm:constant} is straightforward: since actions are differentiated only through their costs, changing $D(a)$ amounts to relabeling actions without affecting equilibrium costs.\footnote{Given any cost function $C(a,\theta)$, equilibrium costs and waste will be unchanged by relabeling actions. Multiplicative costs permit the relabeling to be interpreted as a type-independent change in the cost of an action---the action's ``difficulty''.} The irrelevance to stakes $s$ follows from a two-step decomposition. Scaling both stakes and difficulty by $\alpha>0$ yields a strategically equivalent game, preserving both equilibrium actions and the waste ratio. Scaling difficulty (but not stakes) back down by $1/\alpha$ then leaves equilibrium costs, and hence the waste ratio, unchanged. 

The theorem's second part provides a remarkably simple formula for how much surplus is wasted by signaling in isoelastic environments. The textbook example \citep[e.g.,][p.~329]{FT91} 
with $B(\theta)=\theta$ and $C(a,\theta)=a/\theta$ corresponds to $\beta=\sigma=1$, and so
precisely $50\%$ of the surplus is dissipated.%
\footnote{\label{fn:hms-translation}This waste ratio of $1/2$ also appears in \citet*[Proposition 8]{HoppeMoldovanuSela09}. Translating their two-sided signaling model into our notation agent by agent, and assuming symmetric type distributions, an agent of type $\theta$ who takes signaling action $a$ and is thought to be type $\hat \theta$ receives payoff $\theta \cdot \hat \theta  - a$; maximizing that expression is equivalent to maximizing $\hat \theta - a / \theta$, i.e., a case of $\sigma=\beta$.}
More generally, only the ratio $\beta/\sigma$ matters; waste is monotonically increasing in $\beta/\sigma$, ranging all the way from $0$ to $1$. These directional effects are intuitive. Higher $\beta$ means a greater incentive to separate from lower types; the rat race for higher beliefs becomes fiercer and more of the surplus is burned. Conversely, higher $\sigma$ confers a stronger relative cost advantage to higher types, so separation requires less waste. 

To explain why isoelasticity delivers a constant waste, we present the theorem's proof.

\begin{proof}[Proof of \autoref{thm:constant}]
Substituting $C_a(a,\theta)=D'(a) S(\theta)$ and $V(\theta) = s B(\theta)$ into \autoref{e:foc}, the separating strategy $A$ satisfies (for interior $\theta$) the differential equation
\[
D'(A(\theta)) A'(\theta) = \frac{s B'(\theta)}{S(\theta)}.
\]
As the left-hand side is $\frac{d}{d\theta} D(A(\theta))$, integrate from $0$ to $\theta$ to obtain
\[
D(A(\theta))= s\int_0^\theta \frac{B'(t)}{S(t)} \, dt,
\]
using $D(A(0))=D(0)=0$; when the type space includes $\typemax$, the equality extends to $\theta=\typemax$ by continuity of both sides.
\needspace{3\baselineskip}%
Thus, equilibrium costs are\footnote{Equilibrium costs act similarly to payments in mechanism design, and the derivation of \autoref{e:eqm-cost} is akin to that of the payment identity there \citep{Myerson81}. \autoref{sec:all-pay} fleshes out the link: after a type-by-type rescaling, \eqref{e:eqm-cost} \emph{is} the mechanism-design payment identity, and the signaling game maps to an all-pay auction through which revenue equivalence and order statistics recover the constant-waste formula \eqref{e:constant} under isoelasticity.}
\begin{equation}
\label{e:eqm-cost}
C(A(\theta), \theta) = D(A(\theta)) S(\theta)=s S(\theta) \int_0^\theta \frac{B'(t)}{S(t)} \, dt,
\end{equation}
and the waste ratio is
\begin{equation}
\label{e:invariance}
W(\theta) = \frac{C(A(\theta),\theta)}{V(\theta)} = \frac{S(\theta)}{B(\theta)} \int_0^\theta \frac{B'(t)}{S(t)} \, dt.
\end{equation}
Both the stakes $s$ and the difficulty $D(\cdot)$ have canceled, establishing part \ref{invariance}.

To better interpret the formula \eqref{e:invariance}, define for $\theta>0$ and $t\in [\typemin,\theta]$, 
\[G_\theta(t) := \frac{B(t)/S(t)}{B(\theta)/S(\theta)}.\]
Then $G_\theta$ is a cumulative distribution function on $[\typemin,\theta]$ and we can rewrite \eqref{e:invariance} as%
\footnote{\label{fn:weighted-avg}In more detail: observe that $G_\theta(\typemin)=0$ (using $B(\typemin)=0$),  $G_\theta(\theta)=1$, and for interior $t$, we have $B(t)>0$, $B'(t)> 0$, and $S'(t)<0$, so $(B/S)'(t)>0$.  Hence, $G_\theta$ is strictly increasing on $[\typemin,\theta]$ and is a cumulative distribution function. Its density is $g_\theta(t) = \frac{(B/S)'(t)}{B(\theta)/S(\theta)}$. Rewriting the integrand of \eqref{e:invariance} using $\frac{B'(t)}{S(t)} = \frac{\elaB(t)}{\elaB(t)+\elaS(t)} \cdot \left(\frac{B}{S}\right)'(t)$, which can be verified by expanding $(B/S)'$, and then multiplying by ${S(\theta)}/{B(\theta)}$ yields the integrand $\frac{\elaB(t)}{\elaB(t)+\elaS(t)}g_\theta(t)$.
}
\begin{equation}
\label{e:weighted-avg}
W(\theta) = \int_0^\theta \frac{\elaB(t)}{\elaB(t)+\elaS(t)} \, dG_\theta(t).
\end{equation}
That is, the waste at $\theta$ is a weighted average of the ratios ${\elaB(t)}/{(\elaB(t)+\elaS(t))}$ across types $t<\theta$.

Part \ref{constant} of the theorem follows immediately because in an isoelastic environment the integrand in \eqref{e:weighted-avg} is the constant $\beta/(\beta+\sigma)$. \qedhere

\end{proof}

We highlight that, even outside isoelastic environments,  \autoref{e:weighted-avg} expresses waste at any type $\theta$ as a weighted average of the ratios $\elaB(t)/(\elaB(t)+\elaS(t))$ across all lower types $t<\theta$. We rely on this formula in later analyses.

\begin{example}
\label{eg:isoelastic}
An isoelastic environment with difficulty $D(a)=d \cdot a^\gamma$ for $d>0$ and $\gamma>0$ yields the following separating equilibrium quantities:
    \[A(\theta)=\left(\frac{s \beta}{d (\beta+\sigma)}\right)^{1 / \gamma} \theta^{(\beta+\sigma) / \gamma} \quad \text{and} \quad 
    C(A(\theta), \theta)=\frac{s \beta}{\beta+\sigma} \theta^\beta.\]
Recall that the benefit function is $V(\theta) = s \theta^\beta$ and waste is $W(\theta) = C(A(\theta),\theta)/V(\theta)$. Hence, $W(\theta)=\beta / (\beta+ \sigma)$. So the difficulty parameters $d$ and $\gamma$ affect equilibrium actions, but not costs or benefits, and hence not waste. Stakes $s$ affect actions, costs, and benefits, but not waste.
\end{example}

Multiplicative separability of costs is important for \autoref{thm:constant} part \ref{invariance}; 
\autoref{sec:wasteandstakes} confirms that more generally the waste ratio can either decrease or increase in stakes.\footnote{Multiplicative costs are immaterial for another invariance: the  waste ratio $W(\theta)$ does not depend on the type distribution $F$. This invariance owes to the well-known property that the separating equilibrium strategy only depends on the support of $F$. The strategy discontinuity at complete information carries over to waste; in particular, under isoelasticity, waste equals $\beta/(\beta+\sigma)$ for any full-support $F$, even though it would be zero under complete information.
} Similarly, the isoelastic environment is important for part \ref{constant} of the theorem. In fact, up to a normalization of types, the constant-waste property characterizes isoelasticity under multiplicative costs. That is the content of our next result, whose proof is in \appendixref{sec:charproof}.

\begin{theorem}
\label{thm:char}
Under multiplicative costs, the waste ratio $W(\theta)$ is constant in $\theta > \typemin$ if and only if the ratio of benefit-to-cost elasticities
$\elaB(\theta)/\elaS(\theta)=\rho$ 
for some constant $\rho > 0$. The waste ratio then is $W(\theta)=\elaB(\theta)/(\elaB(\theta)+\elaS(\theta))=
\rho/(1 + \rho)$.
\end{theorem}
 
Thus, when the ratio of the benefit-to-cost elasticities is constant across types, the tension between signaling incentives and costs resolves identically for all types. 
In fact, using \autoref{e:weighted-avg}, \autoref{prop:waste_monotone} in \appendixref{sec:charproof} establishes a more general result: the waste ratio is monotone in type if the benefit-to-cost elasticity ratio is monotone in type.

To see why \autoref{thm:char} characterizes the isoelastic environment up to relabeling types, note that a constant ratio of benefit-to-cost elasticities $\rho$ means $d\ln B/(-d\ln S)=\rho$, and hence 
$S=\kappa B^{-1/\rho}$ for some $\kappa>0$. Since $V$ is strictly  increasing and $V(\typemin)=0$, it can be reparameterized as  $V(t)=st^\beta$ via the change of variable $t=(V(\theta)/s)^{1/\beta}$. 
It follows that $B(t)=t^\beta$, and therefore 
$S(t)=\kappa t^{-\beta/\rho}=\kappa t^{-\sigma}$, where 
$\sigma=\beta/\rho$. Absorbing the constant $\kappa$ into $D(\cdot)$ yields the isoelastic form.

We note that the assumption of multiplicative costs cannot be dropped from \autoref{thm:char}. \autoref{prop:constant-waste-moregeneral} in \autoref{sec:wasteandstakes} shows that a non-multiplicative cost can have constant waste 
without a constant-elasticity structure.

\subsection{Beyond Isoelasticity}
\label{subsec:generalCS}

\autoref{e:constant} implies that waste in isoelastic environments increases when $\beta$ is higher or $\sigma$ is lower. Our next result generalizes that comparative static to non-isoelastic cases.

\begin{theorem}
\label{prop:comp_stat}
Assume multiplicative costs. Consider two environments with benefit and strain elasticities $(\elaB_1,\elaS_1)$ and $(\elaB_2,\elaS_2)$ respectively, and corresponding waste ratios $W_1$ and $W_2$. For any $\theta'>0$, if $\elaB_2(\theta)\geq\elaB_1(\theta)$ and $\elaS_2(\theta)\leq\elaS_1(\theta)$ for all $\theta\in(0,\theta')$, then $W_2(\theta') \geq W_1(\theta')$.
\end{theorem}

Note that $\elaB_2\geq\elaB_1$ is equivalent to $(\ln B_2)'\geq(\ln B_1)'$, which in turn is equivalent to $B_2/B_1$ nondecreasing;  similarly, $\elaS_2\leq\elaS_1$ is equivalent to $(\ln S_2)' \geq (\ln S_1)'$,  or $S_2/S_1$ nondecreasing. So the conditions in \autoref{prop:comp_stat} can also be viewed as monotone ratios.

The intuition for the result can be seen from \autoref{e:weighted-avg}: a pointwise increase in $\elaB$ and decrease in $\elaS$ raises the integrand $\elaB/(\elaB+\elaS)$ pointwise, pushing waste up. The formal proof in \appendixref{sec:proof:compstat} is more nuanced because the weighting distribution $G_\theta$ also differs across environments.\footnote{The proof also establishes that if, in addition, either of the elasticity hypotheses holds strictly on a positive measure of types below $\theta'$, then $W_2(\theta')>W_1(\theta')$.}

\begin{example}
\label{eg:comp_stat}
Consider two environments, one with benefit function $B_1(\theta)=\theta$ and the other with $B_2(\theta)=e^\theta-1$. Both have a common strain function $S(\theta)=\theta^{-1}$. The first environment is thus isoelastic with $\beta = \sigma=1$, and has constant waste $W_1=1/2$. The benefit elasticity in the second environment is $\elaB_2(\theta)=\theta e^\theta/(e^\theta-1) > 1=\elaB_1(\theta)$ for all $\theta>0$.
A straightforward computation from \autoref{e:invariance} yields
\[W_2(\theta)=\frac{1 / \theta}{e^\theta-1} \int_0^\theta t e^t \, d t=\frac{(\theta-1)e^\theta+1}{\theta(e^\theta-1)},\]
which increases from $1/2$ (as $\theta\to 0$) to $1$ (as $\theta\to\infty$). The higher benefit elasticity in the second environment thus yields higher waste at every type.
\end{example}

The elasticity-ranking hypotheses in \autoref{prop:comp_stat} are not necessary for its conclusion; after all, in isoelastic environments, waste only depends on the ratio $\beta/\sigma$. However, the hypotheses are tight in the following sense: if $\elaB_2 (\theta)< \elaB_1(\theta)$ for some type $\theta$, there is a common strain function $S$---in fact, an isoelastic one---such that $W_2 (\theta')< W_1(\theta')$ at some type $\theta'$, and symmetrically for the $\elaS$ condition. See \appendixref{sec:proof:compstat}.

Lastly, there are simple bounds on waste even without knowing exact forms of the benefit or strain functions. Specifically, \autoref{e:weighted-avg} directly yields that if there are constants $\overline \beta$ and $\underline \sigma$ such that $\elaB(\theta) \leq \overline{\beta}$ and $\elaS(\theta) \geq \underline{\sigma}$ for all $\theta$, then $W(\theta) \leq \overline{\beta}/(\overline{\beta}+\underline{\sigma})$ for all $\theta$. Symmetrically, if $\elaB(\theta) \geq \underline{\beta}$ and $\elaS(\theta) \leq \overline{\sigma}$ for all $\theta$, then $W(\theta) \geq \underline{\beta}/(\underline{\beta}+\overline{\sigma})$ for all $\theta$.  

\begin{example}
Suppose the benefit function is $V(\theta) = \theta^2$ (so $\elaB(\theta) = 2$) but the cost function is only known to have strain elasticity $\elaS(\theta) \in [1,3]$ for all $\theta$. Then 
$W(\theta) \in [2/5,\, 2/3]$; in other words, signaling dissipates between 40\% and 67\% of each type's surplus, regardless of the exact strain elasticity.
\end{example}

\section{Is the Waste Worth It?}
\label{sec:welfare}

Our waste ratio only factors in the cost of signaling relative to agents' private benefits from market beliefs. 
The social value of learning agents' types may be different from agents' private value---higher, lower, or even zero. In this section, we explore how the waste from signaling relates to the social value of information in a simple extension of our isoelastic specification. 

Assume that agents are workers in a labor market in which firms compete profits down to zero; hence, the social value of information will be fully internalized into agent payoffs.\footnote{This exercise, comparing payoffs in a regime with no information to another regime with costs of generating information as well as allocative benefits of using it, has analogs in other contexts in \cite{hartline2008optimal}, \citet*[Sections 6--7]{HoppeMoldovanuSela09}, \cite{bulow2012regulated}, and \cite{chakravarty2013optimal}.} 
Specifically, after observing the agent's signaling action, multiple firms each offer a wage to hire the agent. The agent accepts the highest wage offer, and the hiring firm then takes a decision $x\in \Rpos$, which we interpret as an investment level, that yields gross profit
\begin{equation}
\label{e:profit}
\alpha \theta x^\gamma - x.
\end{equation}
So the investment of $x$ is complementary to the agent's type, leading to output  $\alpha \theta x^\gamma$. We assume that $\alpha>0$ and $\gamma \in (0,1)$.\footnote{The linearity of \eqref{e:profit} in $\theta$ is a normalization within the class of power functions. If output were instead $\alpha \theta^r x^\gamma$ for some $r > 0$, we could redefine the type as $\breve \theta = \theta^r$, with corresponding benefit and cost elasticities (discussed subsequently) $\breve \beta = \beta $ and $\breve \sigma = \sigma/r$.}
A routine calculation shows that when a firm believes the agent's type has expectation $\hat \theta$, it expects a profit of $s \hat \theta^\beta$, where $s>0$ depends on $\alpha$ and $\gamma$, and $\beta:=\frac{1}{1-\gamma}>1$.\footnote{Given expected type $\hat \theta$, the firm's optimal investment is $x=(\alpha \gamma \hat \theta)^{\frac{1}{1-\gamma}}$. Substituting into \eqref{e:profit} yields expected profit $\alpha^{\frac{1}{1-\gamma}}\left(\gamma^{\frac{\gamma}{1-\gamma}}-\gamma^{\frac{1}{1-\gamma}}\right) \hat \theta^\beta$.} We assume $\E[\theta^\beta]$ is finite.
 
Thus, firm competition implies the agent's benefit from signaling is $V(\hat\theta) = s \hat\theta^{\,\beta}$. This microfounds our isoelastic benefit specification. The less concave are firms' outputs (higher $\gamma$), the more responsive are their optimal investments and expected profits to the agent's perceived type, and thus the more convex is the agent's benefit function.

Define the {gross separation value} (GSV) as the ex-ante expected social surplus under complete information, the {net separation value} (NSV) as that minus the expected cost of signaling, and the {pooling value} (PV) as the surplus when firms invest based on the agent's mean type. That is,
$$GSV:=s \E[\theta^\beta], \quad NSV:=GSV-\E[C(A(\theta),\theta)], \quad PV:= s (\E[\theta])^\beta.$$

With an isoelastic signaling cost, our waste formula \eqref{e:constant} implies $\mathit{NSV} = \frac{\sigma}{\beta+\sigma}{GSV}$. Consequently, separation has higher net value than pooling if and only if
\begin{equation}
\label{eq:welfare}
\frac{\sigma}{\beta + \sigma} > \frac{(\E[\theta])^\beta}{\E[\theta^\beta]}.
\end{equation}
The left-hand side is the fraction of surplus not dissipated by signaling. The right-hand side---the \emph{pooling efficiency ratio}---is \textit{PV/GSV}, the fraction of surplus captured in a pooling equilibrium (which has no signaling waste). Since $\beta>1$, this pooling efficiency ratio is strictly below $1$ (by Jensen's inequality) and it decreases in $\beta$.\footnote{For $\beta > 1$, Lyapunov's inequality implies $\left(\E[\theta^\beta]\right)^{1 / \beta}$ is increasing in $\beta$, and hence so is $\phi(\beta):=\frac{1}{\beta} \ln \E[\theta^\beta]$. Since
\[
\ln \frac{(\E[\theta])^\beta}{\E[\theta^\beta]}=-\beta(\phi(\beta)-\phi(1)),
\]
it follows that the left-hand side of this display---and hence also the pooling efficiency ratio---is decreasing in $\beta$.}

Inequality \eqref{eq:welfare} is instructive. A higher strain elasticity $\sigma$ reduces signaling waste and unambiguously favors separation. A higher benefit elasticity $\beta$ (stemming from a less concave firm output function) increases the waste ratio but also reduces the pooling efficiency ratio, as the social value of information is larger. The net effect on inequality \eqref{eq:welfare} depends on the type distribution, which affects the pooling efficiency ratio but not the waste ratio. 

We can also look at the role of the type distribution. The distribution does not affect signaling payoffs on the left-hand side of \eqref{eq:welfare}, as previously established, but it does affect pooling efficiency on the right-hand side. Intuitively, information is more valuable when there is more  uncertainty, i.e., when the prior distribution is more spread out. 
Indeed, since $\theta^\beta$ is convex, a mean-preserving spread of types raises $\E[\theta^\beta]$ without changing $(\E[\theta])^\beta$; the pooling efficiency ratio thus falls, favoring separation.
If the support of the type distribution is unbounded ($\typemax=\infty$), then for any fixed $\beta$ and $\sigma$, sufficient type heterogeneity justifies separation regardless of how large the waste ratio is. Conversely, if type heterogeneity is sufficiently limited, then pooling dominates separation. For instance, when $\beta=2$, inequality \eqref{eq:welfare} reduces to the coefficient of variation of $\theta$ exceeding $\sqrt{2/\sigma}$; the type distribution matters only through that one statistic. When $\sigma=2$ as well, the threshold is $1$, which is indeed the condition in \citet*[Proposition 9]{HoppeMoldovanuSela09} for assortative matching to be welfare superior to random matching---their analogs of our separation and pooling.\footnote{This is not a coincidence: following \autoref{fn:hms-translation}, their setting with symmetric type distributions can be translated into ours with $\sigma=\beta$, so that half of each agent's gross benefit from matching is dissipated. Since their multiplicative match output makes that gross benefit quadratic in own type, their comparison also reduces to $(1/2)\E[\theta^2]\gtrless(\E[\theta])^2$, which corresponds to inequality \eqref{eq:welfare} at $\beta=\sigma=2$.}

\begin{example}
Consider a log-normal distribution of types: $\ln\theta \sim \mathcal N (0, \nu^2)$, where the zero mean is a normalization and the variance is $\nu^2>0$.  Here $
\E[\theta^\beta]=\exp\left(\beta^2 \nu^2 / 2\right)$ and $
(\E[\theta])^\beta=\exp\left({\beta \nu^2 / 2}\right)$, so the pooling efficiency ratio is $$\exp(-\nu^2\beta(\beta-1)/2).$$ Pooling efficiency thus decays exponentially in both the variance of log-types and the benefit elasticity. Rearranging \eqref{eq:welfare}, separation dominates pooling despite signaling's waste if and only if
\begin{align} 
\label{eq:lognormal}
\nu^2 > \frac{2\ln\left(1 + {\beta}/{\sigma}\right)}{\beta(\beta-1)}.
\end{align}
A higher variance $\nu^2$ makes separation relatively more appealing than pooling. 
In this example, a higher benefit elasticity $\beta$ also makes separation relatively more appealing, as one can verify that the right-hand side of \eqref{eq:lognormal} decreases in $\beta$.%
\end{example}

\section{Application: Competition and Waste}
\label{sec:tournament}

We now apply our results to signaling tournaments, in which agents compete for a prize. Specifically, a set of agents each draw a type and take a signaling action. After observing these actions,  a designer awards a prize---a job offer, a promotion, college admission---to the agent with the highest inferred type. We will use our earlier results to show that more agents or fewer prizes increase signaling waste. In other words, more competitive tournaments yield more waste.

\subsection{Symmetric Agents and One Prize}
\label{sec:tournament-simpler}

To begin, consider a tournament among $N\geq 2$ agents for a single prize of value $s>0$.  Agents' types are their private information, drawn independently from a common cumulative distribution $F$ on $[0,1]$ with continuous density $f(\theta)>0$ for $\theta>0$. Agents simultaneously choose their observable signaling actions $a$, at multiplicative cost $C(a,\theta)$ satisfying \autoref{ass:basics} part \ref{ass:C}. The prize is awarded to the agent with the highest perceived type.

If agent $i$ is perceived as type $\hat\theta_i$, her probability of winning is $\Prob(\hat\theta_i > \max_{j \neq i} \hat\theta_j)$. In a symmetric separating equilibrium, an agent's expected benefit is thus
\[
V(\theta) = s \cdot \left(F(\theta)\right)^{N-1}.
\]
So the tournament induces a signaling game with a specific benefit function that depends on the type distribution and the number of agents.
Letting $B(\theta):=\left(F(\theta)\right)^{N-1}$, the benefit elasticity $\elaB$ is
\[
\elaB(\theta)  = \theta\frac{B'(\theta)}{B(\theta)}= \theta \frac{(N-1) f(\theta) \left(F(\theta)\right)^{N-2}}{\left(F(\theta)\right)^{N-1}} = (N-1) \frac{   \theta f(\theta)}{F(\theta)}.
\]

An immediate observation is that the benefit elasticity at each type $\theta$ is increasing in the number of agents $N$. Hence, by \autoref{prop:comp_stat}, the waste ratio is increasing in the number of agents. Moreover, $\elaB(\theta)$ grows without bound in $N$ at each $\theta>0$, so the integrand $\elaB(t)/(\elaB(t)+\elaS(t))$ in \eqref{e:weighted-avg} tends to one pointwise. \autoref{prop:tournament} below confirms that all surplus is dissipated in large tournaments as $N\to \infty$.

With further assumptions on the type distribution and cost function, we obtain an isoelastic environment. Consider a power-function distribution $F(\theta) = \theta^z$ for some $z>0$. Then the benefit function is $B(\theta) = \theta^{(N-1) z}$, with constant benefit elasticity $\beta = (N-1)z$. If costs are also of the isoelastic form $C(a,\theta) = D(a) \theta ^{-\sigma}$ with constant strain elasticity $\sigma>0$, then
\autoref{thm:constant} delivers a constant waste ratio of
\begin{equation}\label{e:tournament}
W(\theta) = \frac{(N-1)z}{(N-1)z + \sigma }.
\end{equation}
In particular, for a uniform distribution ($z=1$) and unit strain elasticity ($\sigma=1$), the constant waste is simply $(N-1)/N$. More generally, in the isoelastic case waste increases in the type distribution's elasticity $z$, because a higher $z$ translates into greater convexity of the benefit function. Waste also decreases in the cost strain elasticity $\sigma$, as in our general analysis.

\subsection{Multiple Prizes and Asymmetric Agents}

The above logic can be extended to tournaments with multiple prizes and with heterogeneous agents.

Specifically, consider $m$ identical prizes, where $1 \le m<N$. Prizes will be awarded to the $m$ agents with the highest inferred types. 
Each agent $i$ has a multiplicative cost function $C_i(a,\theta)$, value $s_i$ of winning a prize, and type independently drawn from a cumulative distribution $F_i$ on $[0,1]$ with continuous density $f_i(\theta)>0$ for $\theta>0$. The parameters $C_i$, $s_i$, and $F_i$ are all commonly known; an agent's private information is only her realized type $\theta_i$. We say agents are \emph{symmetric} if the parameters $(C_i, s_i, F_i)$ do not vary with $i$. 

In a separating equilibrium, agent $i$'s value $V_i(\theta)$ is $s_i$ times the probability that at most $m-1$ others have types above $\theta$. Because of their heterogeneity, agents may have different separating strategies; agent $i$'s separating strategy $A_i(\theta_i)$ depends on her own cost $C_i$ and prize value $s_i$, and on her competitors' type distributions $(F_{j})_{j\neq i}$.\footnote{As elaborated in \autoref{sec:all-pay}, with symmetric agents the separating equilibrium of the signaling tournament is equivalent to the symmetric equilibrium of a standard all-pay auction with $m$ identical objects; an agent's private type determines her bidding cost rather than her value of winning, but that is isomorphic. Heterogeneity then yields an asymmetric all-pay auction with bidder-specific rules for winning. The rule that awards the prizes to the top $m$ inferred types can be thought of as the ex post efficient rule when social welfare is maximized by allocating prizes to higher types.}

We fix a universe of possible agents, $\{1,2,\ldots\}$, with each agent $i$ described by $(C_i, s_i, F_i)$, and consider the signaling tournament with $m$ prizes and agents $1,\ldots,N$. Let the waste ratio for agent $i\le N$ in the corresponding separating equilibrium be given by $W_i^{m,N}(\theta)$.

\begin{proposition} \label{prop:tournament}
In a signaling tournament with $m$ prizes and $N$ agents, any agent $i$'s waste ratio $W_i^{m,N}(\theta)$ is increasing in $N$ and decreasing in $m$ at every type $\theta>0$. Furthermore, if agents are symmetric, then for any $\theta>0$, we have $\lim\limits_{N\to \infty} W_i^{m,N}(\theta)=1$.
\end{proposition}

In other words, a more competitive tournament---either because of more agents or fewer prizes---leads to more waste.  These comparative statics follow from \autoref{prop:comp_stat} once we establish (in \appendixref{sec:proof:tournament}) that higher $N$ and lower $m$ both imply pointwise increases in each agent $i$'s benefit elasticity $\elaB_i(\theta)$. Moreover, with symmetric agents and a fixed number of prizes, eventually every type of every agent dissipates all her surplus.\footnote{In fact, full dissipation admits a characterization beyond symmetric agents. Holding agent $i$'s cost function fixed, consider any sequence of tournaments in which the other agents, the type distributions, and the number of prizes may change with $N$. Then agent $i$'s waste ratio at $\theta>0$ converges to one if and only if every type $t<\theta$ is eventually shut out of contention relative to $\theta$: the ratio of agent $i$'s probability of winning a prize when her type is $t$ to that probability when her type is $\theta$ tends to zero. See \autoref{rem:shutout} in \appendixref{sec:proof:tournament}.} There is also full dissipation in the aggregate: even though each agent's expected signaling cost vanishes as the field grows, expected total costs across the $N$ agents converge to $s\cdot m$, the full value of the prizes (\autoref{prop:aggregate-dissipation}). 

\subsection{Discussion}
The above analysis gives a different take on contests and competition than canonical \citet{Tullock80} contests (see \citet{Nitzan94} and \citet{Corchon07} for surveys). In Tullock contests, the vector of actions stochastically determines a winner. For instance, if symmetric agents pay an isoelastic cost $x_i^\gamma$ (with $\gamma\ge 1$) to win with probability $x_i / \sum_j x_j$, or equivalently pay a cost of $x_i$ to win with probability $x_i^{1/\gamma}/ \sum_j (x_j)^{1/\gamma}$, the rent dissipation rate is $\frac{1}{\gamma}\frac{N-1}{ N}$.

Our signaling tournament is more like an all-pay auction, where the actions deterministically yield a winner, albeit through an endogenous mapping. Winning is uncertain only because an agent does not know rivals' types, and thus their actions. Unlike in a Tullock contest with strictly convex costs $(\gamma>1)$, a growing field of symmetric agents drives waste to full dissipation in the limit. Intuitively, with strictly convex costs, the noise in a Tullock contest can preserve some surplus even under extreme competition, whereas separating from a dense field of rivals forces full dissipation under signaling. The waste ratios in the two models coincide for every $N$ only when the Tullock cost is linear ($\gamma=1$) and the signaling tournament's strain elasticity matches the type distribution's elasticity ($\sigma=z$).

More generally, the two models differ in how waste responds to the cost of effort. The Tullock rate $\frac{1}{\gamma}\frac{N-1}{N}$ falls in the cost convexity $\gamma$: steeper marginal costs discourage effort. By contrast, \autoref{thm:constant} renders the difficulty $D(\cdot)$ irrelevant to signaling waste. Separation forces agents to scale their actions with $D(\cdot)$ exactly enough to leave the waste ratio unchanged. Signaling waste thus cannot be reduced by making better performance costlier; it falls only with a lower benefit elasticity (fewer competitors or more prizes) or a higher strain elasticity.

Tournament-like signaling and matching have also been studied in richer models by \citet{Hopkins12} and, as discussed earlier, \citet*{HoppeMoldovanuSela09}. That the latter's dissipation is bounded by one half even in large markets---whereas our one-sided tournament dissipates the entire surplus as $N\to\infty$---reflects their two-sided structure: each agent dissipates only the own-side rent from winning a better partner, while the other half of the surplus accrues to that partner. In our tournament, no partner absorbs surplus, and competition instead drives waste toward one.

\section{Conclusion}
\label{sec:conclusion}

We have proposed the \hyperref[def:waste]{waste ratio}---the fraction of a type $\theta$'s surplus dissipated through signaling---as a natural measure of signaling inefficiency in separating equilibria. 
Our main results tie waste to two elasticities under multiplicative costs: the benefit elasticity $\elaB(\theta)$ and the strain elasticity $\elaS(\theta)$, a measure of higher types' comparative advantage.  \autoref{thm:constant} shows that the waste ratio is invariant to signaling stakes and difficulty, and in isoelastic environments where $\elaB(\theta)\equiv\beta$ and $\elaS(\theta)\equiv\sigma$ it is the constant $\beta/(\beta+\sigma)$. More generally, \autoref{prop:comp_stat} shows that waste rises at every type when the benefit elasticity rises pointwise or the strain elasticity falls pointwise.

The invariance to difficulty undermines some common intuitions about the inefficiency of signaling. Consider the debate about the difficulty of standardized tests for college admissions. Recent trends favor shorter, less complex tests (such as the digital SAT) to reduce student stress. 
Conversely, some critics call for harder exams to restore selectivity. Our results suggest that, when viewed through the canonical signaling lens, neither approach may address the underlying waste. Adjusting the difficulty of the test uniformly for all students---whether making it easier or harder---need not change the total resource dissipation; it could merely rescale equilibrium effort while leaving the waste ratio constant.\footnote{This invariance does rely on a single-dimensional framework; it could break if students also differ in test-taking aptitude separate from underlying ability, which would lead to ``muddled information'' \citep{FK19}.} As long as admissions at selective colleges resemble winner-take-all signaling tournaments with many competitors, the process is likely to dissipate significant surplus, regardless of how the testing technology is calibrated.

Of course, college admission itself is not the final prize. There are concerns about the ``winner-take-all'' nature of the broader society \citep[e.g.,][]{frank1996winner}. For a fixed distribution of underlying types, our isoelastic specification captures the inequality of socioeconomic outcomes via the benefit elasticity $\beta$. In particular, with value function $V(\theta) = s \theta^\beta$, a higher $\beta$ corresponds to more inequality via a more convex mapping from types to benefits. Our results show that a higher $\beta$---corresponding, perhaps, to the US versus lower inequality countries like Canada or Sweden---goes hand in hand with more waste from signaling.

The strain elasticity $\sigma$ also matters. Another approach to reducing waste would be to make signaling instruments more discriminating in the sense of increasing $\sigma$. Returning to exams, a redesigned test that amplifies high-ability students' comparative advantage---rather than scaling difficulty uniformly---would correspond to increasing $\sigma$ and would indeed reduce waste. Interestingly, this echoes a discussion in the biological signaling literature: animals can often reliably convey information while incurring minimal waste. The mechanism stems from sharply different marginal costs across types---originally proposed by \citet{Zahavi77} to refine his earlier ``handicap'' hypothesis---rather than difficulty. Scholars have argued that because Darwinian selection favors efficiency, it leads to biological signals that are cheap for high-quality types but prohibitive for low-quality types \citep{PennSzamado20}, corresponding to a high strain elasticity $\sigma$.

We close by commenting on some limitations of our analysis. First, we only study separating equilibria. That is consistent with much of the literature's emphasis, often justified by stability-based arguments \citep[e.g.,][]{ChoSobel90}. But there are also critiques of the exclusive focus on separating equilibria \citep[e.g.,][]{MOFP93}. Equilibria with some pooling, where certain types choose identical actions, can reduce signaling costs and waste \citep[e.g.,][]{krishna2026pareto}.

Second, we focus on the dissipative cost of signaling activities. Such activities can, of course, sometimes be productive; see \autoref{fn:productive}. We expect that even in a broader welfare calculus, our waste ratio is a useful input: the net social value must weigh signaling's intrinsic benefit against its waste.

Lastly, outside the isoelastic class, we have comparative statics (\autoref{prop:comp_stat}) and a generalization of the isoelastic formula (\autoref{e:weighted-avg}), but ultimately no simple expression for the waste ratio. Moreover, \autoref{thm:char} indicates that the waste ratio will generally vary with type, meaning that aggregate waste will depend on the type distribution. We do not suggest that isoelasticity should be taken literally. Rather, we view it as focal by analogy to how CRRA utility is canonical not because preferences literally exhibit constant relative risk aversion, but because it affords tractable analyses and scale-free results. We hope the waste formula $\beta/(\beta + \sigma)$ is a similarly useful benchmark for signaling's welfare cost. 

\appendix

\section{Separating Equilibria}
\label{sec:separating}

A \emph{(mixed) strategy} for the agent is a measurable mapping $\alpha: \Theta \to \Delta(\Rpos)$, where $\Delta(\Rpos)$ denotes the set of probability distributions over actions. A \emph{pure strategy} is a strategy $\alpha$ such that $\alpha(\theta)$ has singleton support for all $\theta$; we denote a pure strategy more simply by $A:\Theta \to \Rpos$.
Since a belief concentrated on type $\typemin$ is the ``worst belief'' (by monotonicity of the benefit function $V$), and hence is the most conducive off-path belief to support an equilibrium, we say that strategy $\alpha$ defines a \emph{separating equilibrium} if there is a measurable belief map $\hat\theta: \Rpos \to \Theta$ such that:
\begin{enumerate}
    \item \label{sepdef1} (Separation.) For every $\theta \in \Theta$, we have $\hat\theta(a) = \theta$ for $\alpha(\theta)$-a.e.~$a$.
    \item \label{sepdef2} (Incentive compatibility.) For each $\theta \in \Theta$, $\alpha(\theta)$-a.e.~$a$, and all $a' \in \Rpos$:
    \[
    V(\theta) - C(a, \theta) \geq V(\hat \theta(a')) - C(a', \theta).
    \]
\end{enumerate}
Without loss, we can set $\hat\theta(a) = \typemin$ at off-path actions.

\begin{proposition}\label{prop:separating-differentiable}
Any separating equilibrium is a pure-strategy equilibrium. Moreover, its strategy $A: \Theta \to \Rpos$ is continuous and strictly increasing on $[\typemin,\typemax\rangle$, 
differentiable on $(\typemin, \typemax)$, and satisfies
\begin{equation}
\label{e:ODE}
A'(\theta) = \frac{V'(\theta)}{C_a(A(\theta), \theta)} \quad \text{for } \theta \in (\typemin,\typemax),
\end{equation}
with boundary condition $A(\typemin) = 0$.
\end{proposition}

Although we are not aware of an existing result that directly implies \autoref{prop:separating-differentiable}, the proof follows familiar lines \citep[cf.~][]{Mailath87} and is provided in the \suppapp{sec:proof:separating-differentiable}.  It is worth noting that because \autoref{prop:separating-differentiable} only establishes necessary conditions for a separating equilibrium, it does not require all of \autoref{ass:basics}; in particular, it is enough that $V$ is differentiable on the interior of the type space (rather than continuously differentiable), that $C_a$ is continuous (rather than differentiable), and that $C(\cdot,\theta)$ is strictly increasing---where finite for type $0$---for all $\theta$ (it is not necessary that $C_{a\theta}<0$). The additional properties are instead used to verify sufficiency in the next result, whose proof---which largely follows \citet{Mailath87}---is also in the \suppapp{sec:proof:separating-differentiable}.

\begin{proposition}
\label{prop:sufficiency}
A continuous function $A: \Theta \to \Rpos$ that is differentiable on $(0,\overline \theta)$ and satisfies \eqref{e:ODE} and $A(0)=0$ constitutes a separating equilibrium.
\end{proposition}

\section{Proof of \autoref{thm:char}}
\label{sec:charproof}

Let $\elaR(\theta):=\elaB(\theta)/\elaS(\theta)$ denote the ratio of benefit-to-cost elasticities. We first prove the following result.

\begin{proposition}
\label{prop:waste_monotone}
Assume multiplicative costs. If $\elaR(\theta)$ is nondecreasing, then $W(\theta)$ is nondecreasing; if $\elaR(\theta)$ is nonincreasing, then $W(\theta)$ is nonincreasing.
\end{proposition}

\begin{proof}
Recall from \autoref{e:weighted-avg} in the proof of \autoref{thm:constant} that
$$W(\theta) = \int_\typemin^\theta \frac{\elaB(t)}{\elaB(t)+\elaS(t)} \, dG_\theta(t).$$
Consider any $\theta_2>\theta_1$. The distribution $G_{\theta_2}$ first-order stochastically dominates $G_{\theta_1}$ because, for $t \leq \theta_1$, the monotonicity of $B/S$ implies
$$G_{\theta_2}(t) = \frac{B(t)/S(t)}{B(\theta_2)/S(\theta_2)} \leq \frac{B(t)/S(t)}{B(\theta_1)/S(\theta_1)} = G_{\theta_1}(t).$$
If $\elaR$ is nondecreasing, so is $\elaB/(\elaB+\elaS)$, and hence $W(\theta_2) \geq W(\theta_1)$. The case of $\elaR$ nonincreasing is symmetric. \qedhere
\end{proof}

\autoref{eg:comp_stat} presented one environment with an increasing ratio of benefit-to-cost elasticities: $\elaR_2(\theta) = \theta e^\theta/(e^\theta - 1)$. The computation there of an increasing waste ratio function $W_2$ illustrates \autoref{prop:waste_monotone}.

\begin{proof}[Proof of \autoref{thm:char}]
If $\elaR(\theta)\equiv \rho$ is constant, \autoref{prop:waste_monotone} implies $W$ is both nondecreasing and nonincreasing, hence constant. The converse argument below then gives $W=\rho/(1+\rho)$. 

Conversely, if $W(\theta)\equiv k\in (0,1)$ is constant, differentiating \autoref{e:invariance} gives $0=(\ln B)'(1-k)+(\ln S)'k$, and hence $\elaR(\theta) = (\ln B)'/(-(\ln S)') = k/(1-k)$, a constant. Setting $\rho:=k/(1-k)$ gives $W=\rho/(1+\rho)$.
\end{proof}

\section{Proof of \autoref{prop:comp_stat}} 
\label{sec:proof:compstat}

We state and prove a result for general (non-multiplicative) costs that implies \autoref{prop:comp_stat}. We present the more general result because, even though it relies on an endogenous object when costs are not multiplicative, it clarifies the underlying logic. 

When costs are not multiplicative, we additionally assume $C(a,\theta)$ is differentiable in $\theta$ with $C_\theta$ continuous; under multiplicative costs, this holds automatically because $C_\theta(a,\theta)=D(a)S'(\theta)$. Given a separating function $A$, define the \emph{on-path cost elasticity} by
\[
\elaS^A(\theta) := -\theta \frac{C_\theta(A(\theta),\theta)}{C(A(\theta),\theta)}.
\]
Under multiplicative costs, this simplifies to the primitive strain elasticity $\elaS(\theta)$ because for any action $a$, it holds that
\begin{equation}
\label{e:onpath-elaS-simplifies}
-\theta \frac{C_\theta(a,\theta)}{C(a,\theta)}
= -\theta \frac{D(a)S'(\theta)}{D(a)S(\theta)}
= -\theta \frac{S'(\theta)}{S(\theta)}.
\end{equation}

\begin{proposition}
\label{prop:comp_stat_nonmult}
Consider two environments with benefit elasticities $\elaB_1$ and $\elaB_2$, 
separating equilibria $A_1$ and $A_2$, on-path cost elasticities $\elaS^{A_1}_1$ and $\elaS^{A_2}_2$, and waste ratios $W_1$ and $W_2$. For any $\theta' > 0$, if $\elaB_2(\theta) \ge \elaB_1(\theta)$ and $\elaS^{A_2}_2(\theta) \le \elaS^{A_1}_1(\theta)$ for all $\theta \in (0,\theta')$, then $W_2(\theta') \ge W_1(\theta')$.
\end{proposition}

This result immediately implies \autoref{prop:comp_stat} because of the simplification \eqref{e:onpath-elaS-simplifies} under multiplicative costs.

\begin{proof}[Proof of \autoref{prop:comp_stat_nonmult}]
For each $i\in\{1,2\}$, let $C^*_i(\theta) := C_i(A_i(\theta),\theta)$, so that $W_i(\theta)=C^*_i(\theta)/V_i(\theta)$. Differentiating $C^*_i$ and using \autoref{e:ODE}, we have
\[
(C^*_i)'(\theta) = \frac{\partial}{\partial a} C_i(A_i(\theta),\theta)A_i'(\theta) + \frac{\partial}{\partial \theta}C_{i}(A_i(\theta),\theta)
= V_i'(\theta) + \frac{\partial}{\partial \theta}C_{i}(A_i(\theta),\theta).
\]
Hence,
\[
W_i'(\theta) = \frac{(C^*_i)'(\theta)}{V_i(\theta)} - \frac{C^*_i(\theta)V_i'(\theta)}{(V_i(\theta))^2}
= \frac{V_i'(\theta)}{V_i(\theta)}(1-W_i(\theta)) + \frac{1}{V_i(\theta)}\,\frac{\partial}{\partial \theta}C_{i}(A_i(\theta),\theta).
\]
Using
\[
\frac{V_i'(\theta)}{V_i(\theta)} = \frac{\elaB_i(\theta)}{\theta}
\quad\text{and}\quad
\frac{1}{V_i(\theta)}\,\frac{\partial}{\partial \theta}C_{i}(A_i(\theta),\theta)
= \frac{W_i(\theta)}{C^*_i}\,\frac{\partial}{\partial \theta}C_{i}(A_i(\theta),\theta)
= -\frac{\elaS_i^{A_i}(\theta)}{\theta}\,W_i(\theta),
\]
we obtain
\begin{equation}
\label{e:general-Wprime}
W_i'(\theta) = \frac{\elaB_i(\theta)}{\theta}\left(1-W_i(\theta)\right) - \frac{\elaS_i^{A_i}(\theta)}{\theta}\,W_i(\theta).
\end{equation}

Let $\Delta(\theta):=W_2(\theta)-W_1(\theta)$. Then
\begin{equation}
\label{e:Delta'_nonmult}
\Delta'(\theta) + \frac{\elaB_1(\theta)+\elaS_1^{A_1}(\theta)}{\theta}\,\Delta(\theta) = \gamma(\theta),
\end{equation}
where
\[
\gamma(\theta)
:=
\frac{\elaB_2(\theta)-\elaB_1(\theta)}{\theta}\left(1-W_2(\theta)\right)
+
\frac{\elaS_1^{A_1}(\theta)-\elaS_2^{A_2}(\theta)}{\theta}\,W_2(\theta).
\]
Since $W_2\in(0,1)$, the elasticity assumptions ensure $\gamma\ge 0$ on $(0,\theta')$.

Multiplying both sides of \eqref{e:Delta'_nonmult} by
\[
H(\theta) := \exp\!\left(-\int_\theta^{\theta'} \frac{\tilde \beta_1(t)+\tilde \sigma_1^{A_1}(t)}{t}\,dt\right)
\]
gives
\[
\frac{d}{d\theta}[\Delta(\theta)H(\theta)] = \gamma(\theta)H(\theta).
\]
Integrating from $0$ to $\theta'$, we have
\[
\Delta(\theta')H(\theta') - \lim_{t\to 0}\Delta(t)H(t) = \int_0^{\theta'} \gamma(t)H(t)\,dt.
\]
Observe that $H(\theta')=1$, and the limit term is zero because $\Delta$ is bounded and, as $t\to 0$, $H(t)\to 0$ because
\[
\int_t^{\theta'} \frac{\tilde \beta_1(u)+\tilde \sigma_1^{A_1}(u)}{u}\,du
\ge \int_t^{\theta'} \frac{\tilde \beta_1(u)}{u}\,du
= \int_t^{\theta'} \frac{V_1'(u)}{V_1(u)}\,du
= \ln V_1(\theta') - \ln V_1(t) \to \infty.
\]
Therefore,
\[
\Delta(\theta') = \int_0^{\theta'} \gamma(t)H(t)\,dt \ge 0. \qedhere
\]
\end{proof}

See \autoref{eg:nonmultsepcost} for an illustration of \autoref{prop:comp_stat_nonmult}.

\begin{proof}[Proof of tightness of the conditions in \autoref{prop:comp_stat}]
We show that if $\elaB_2(\tilde\theta)<\elaB_1(\tilde\theta)$ at some type $\tilde\theta\in(\typemin,\typemax)$, then there exists a common isoelastic strain function $S$ across the two environments such that $W_2(\tilde\theta)<W_1(\tilde\theta)$. The argument for the $\elaS$ condition is analogous.

Consider the family of isoelastic strain functions $S_\sigma(\theta)=\theta^{-\sigma}$ for $\sigma>0$. Evaluating \autoref{e:invariance} at $\tilde\theta$ under $S_\sigma$ gives
\[
W_i(\tilde\theta;\sigma)
=
\frac{\tilde\theta^{-\sigma}}{B_i(\tilde\theta)}
\int_{\typemin}^{\tilde\theta} B_i'(
\theta)\,\theta^\sigma\,d\theta.
\]
Multiplying by $\sigma+1$ and changing variables to $\theta=x\tilde\theta$ yields
\begin{equation}
\label{e:tightness}
(\sigma+1)W_i(\tilde\theta;\sigma)
=
\int_0^1 \frac{\tilde\theta\, B_i'(x\tilde\theta)}{B_i(\tilde\theta)}\,(\sigma+1)x^\sigma\,dx.
\end{equation}
As $(\sigma+1)x^\sigma$ is a density on $[0,1]$ which converges in distribution to the point mass at $x=1$ as $\sigma\to\infty$, and $B_i'$ is continuous on $(0,\tilde\theta]$,\footnote{Although $B_i'$ may be unbounded near $\typemin$ (e.g., when $B_i(\theta)=\theta^\beta$ with $\beta<1$), types near $\typemin$ contribute negligibly to \eqref{e:tightness}: for any fixed $\varepsilon\in(0,1)$, since $x^\sigma\le\varepsilon^\sigma$ on $[0,\varepsilon]$ and $\int_0^\varepsilon \tilde\theta B_i'(x\tilde\theta)\,dx=B_i(\varepsilon\tilde\theta)$, the contribution of $[0,\varepsilon]$ to the integral in \eqref{e:tightness} is at most $(\sigma+1)\varepsilon^\sigma B_i(\varepsilon\tilde\theta)/B_i(\tilde\theta)$, which goes to $0$ as $\sigma\to \infty$. On $[\varepsilon,1]$ the integrand is bounded and continuous.} it follows that
\[
\lim_{\sigma\to\infty}(\sigma+1)\,W_i(\tilde\theta;\sigma)
=
\frac{\tilde\theta\, B_i'(\tilde\theta)}{B_i(\tilde\theta)}
=
\elaB_i(\tilde\theta).
\]
Hence, $\elaB_2(\tilde\theta)<\elaB_1(\tilde\theta)$ implies that for sufficiently large $\sigma$ we have $$(\sigma+1)W_2(\tilde\theta;\sigma)<(\sigma+1)W_1(\tilde\theta;\sigma),$$ or equivalently $W_2(\tilde\theta;\sigma)<W_1(\tilde\theta;\sigma)$.
\end{proof}

\section{Proofs and Additional Results for \autoref{sec:tournament}}
\label{sec:proof:tournament}

\subsection{Proof of \autoref{prop:tournament}}

Fix a focal agent facing a set $J$ of opponents. Let $P_{J,0}:=0$ and for $1\le m \leq |J|$, let $P_{J,m}(\theta)$ denote the probability that the focal agent of type $\theta$ wins one of $m$ identical prizes, i.e., that at most $m-1$ opponents in $J$ have types above $\theta$. The agent's benefit is proportional to $P_{J,m}(\theta)$, so the benefit elasticity is the elasticity of $P_{J,m}$. For the proposition's first statement, it suffices, by \autoref{prop:comp_stat}, to show that this elasticity rises when an opponent is added to $J$ (increasing the number of agents) or when $m$ is lowered (decreasing the number of prizes). For both these comparative statics, we will use the following lemma, whose proof is later in the section.

\begin{lemma}\label{lem:competition}
For any opponent set $J$ and prizes $m\in\{1,\dots,|J|\}$, the ratio $P_{J,m-1}(\theta)/P_{J,m}(\theta)$ is increasing in $\theta$ on $(0,1]$.
\end{lemma}

To interpret the lemma, let $X_\theta$ be the number of opponents in $J$ with types above $\theta$, so that $P_{J,m}(\theta)=\Prob(X_\theta\le m-1)$ and
\[
\frac{P_{J,m-1}(\theta)}{P_{J,m}(\theta)}=\Prob\bigl(X_\theta\le m-2\mid X_\theta\le m-1\bigr).
\]
The lemma thus says that, conditional on winning one of $m$ prizes, a higher type is more likely to have ranked high enough to win one of only $m-1$; intuitively, a higher $\theta$ stochastically lowers the number of opponents above it.

\paragraph{Comparative static in the number of agents.}
Add a new opponent, indexed $0$, with type distribution $F_0$. The focal agent of type $\theta$ then wins one of the $m$ prizes against $J\cup\{0\}$ either if opponent $0$'s type falls below $\theta$ (which has probability $F_0(\theta)$) and the agent would have won against $J$, or if it falls above $\theta$ (probability $1-F_0(\theta)$) and the agent would have won one of $m-1$ prizes against $J$. Hence
\[
P_{J\cup\{0\},m}(\theta)=F_0(\theta)\,P_{J,m}(\theta)+\bigl(1-F_0(\theta)\bigr)P_{J,m-1}(\theta).
\]
Dividing by $P_{J,m}(\theta)$ yields
\[
\frac{P_{J\cup\{0\},m}(\theta)}{P_{J,m}(\theta)}=F_0(\theta)+\bigl(1-F_0(\theta)\bigr)\frac{P_{J,m-1}(\theta)}{P_{J,m}(\theta)}.
\]
The right-hand side is increasing in $\theta$: it is a convex combination of $1$ and $P_{J,m-1}/P_{J,m}\le1$, in which both the weight $F_0$ placed on $1$ and the value $P_{J,m-1}/P_{J,m}$ increase in $\theta$ (the latter by \autoref{lem:competition}). Therefore $\tfrac{d}{d\theta}\bigl[\log P_{J\cup\{0\},m}(\theta)-\log P_{J,m}(\theta)\bigr]\ge0$, so adding an opponent raises the focal agent's benefit elasticity pointwise.

\paragraph{Comparative static in the number of prizes.}
For $m\ge2$, taking logarithms, \autoref{lem:competition} states that $\log P_{J,m-1}(\theta)-\log P_{J,m}(\theta)$ is increasing in $\theta$. Hence reducing the number of prizes from $m$ to $m-1$ raises the focal agent's benefit elasticity pointwise.

\begin{proof}[Proof of \autoref{lem:competition}]
The claim is trivially true when $m=1$, as $P_{J,0}\equiv0$. So fix any $m\ge2$ and any $0<\theta<\theta'\le1$. We will show that $P_{J,m-1}(\theta')/P_{J,m}(\theta')\geq P_{J,m-1}(\theta)/P_{J,m}(\theta)$.

Let $X:=\#\{j\in J:\theta_j>\theta\}$ and $X':=\#\{j\in J:\theta_j>\theta'\}$ count the opponents whose types exceed $\theta$ and $\theta'$ respectively, so that $P_{J,m}(\theta)=\Prob(X\le m-1)$ and $P_{J,m}(\theta')=\Prob(X'\le m-1)$. Each of $X$ and $X'$ is a sum of independent Bernoulli random variables, one for each opponent $j\in J$, with success probabilities $1-F_j(\theta)$ and $1-F_j(\theta')$ respectively. Since $1-F_j(\theta')\le1-F_j(\theta)$ for every $j$, it follows from \citet[Example 1.C.10]{ShakedShanthikumar2007} that $X$ dominates $X'$ in the likelihood-ratio order.\footnote{Their example establishes that for a sum of independent Bernoulli random variables, the ratio of the probability of $k+1$ successes to that of $k$ is increasing in each of the underlying success probabilities. Changing those success probabilities one at a time gives the likelihood-ratio comparison between $X$ and $X'$.} Writing $p_k:=\Prob(X=k)$ and $q_k:=\Prob(X'=k)$, it follows that $q_k/p_k$ is decreasing in $k$; in particular, $q_k/p_k\ge q_{m-1}/p_{m-1}$ for every $k\le m-2$. Summing over these $k$ and rearranging gives
\[
\Bigl(\sum_{k=0}^{m-2} q_k\Bigr)\, p_{m-1}\;\ge\;\Bigl(\sum_{k=0}^{m-2} p_k\Bigr)\, q_{m-1},
\]
which by cross-multiplication and simplification is equivalent to
\[
\frac{P_{J,m-1}(\theta')}{P_{J,m}(\theta')}
=\frac{\sum_{k=0}^{m-2}q_k}{\sum_{k=0}^{m-2}q_k+q_{m-1}}
\;\ge\;
\frac{\sum_{k=0}^{m-2}p_k}{\sum_{k=0}^{m-2}p_k+p_{m-1}}
=\frac{P_{J,m-1}(\theta)}{P_{J,m}(\theta)}.\qedhere
\]
\end{proof}

\paragraph{Full dissipation as the field grows.}
Finally, we prove the limit claim of \autoref{prop:tournament}. Let agents be symmetric, with multiplicative cost $C(a,\theta)$ and type distribution $F$, and fix the number of prizes $m$. Let $P_n(\theta)$ denote $P_{J,m}(\theta)$ for an opponent set $J$ of size $n:=N-1\ge m$, so that $P_n(\theta)$ is the probability that at most $m-1$ of $n$ opponents drawn independently from $F$ have types above $\theta$. The focal agent's benefit is proportional to $P_n$, so by \autoref{e:invariance} her waste ratio is $W_i^{m,N}(\theta)=\frac{S(\theta)}{P_n(\theta)}\int_0^\theta \frac{P_n'(t)}{S(t)}\,dt$. Integrating by parts, using $P_n(0)=0$ and $1/S(t)\le 1/S(\theta)$ to see that the boundary term at $t=0$ vanishes, we have
\begin{equation}\label{e:rent-share}
1-W_i^{m,N}(\theta)=\int_0^\theta \frac{P_n(t)}{P_n(\theta)}\,w(t)\,dt,
\qquad\text{where } w(t):=S(\theta)\Bigl(\frac{1}{S}\Bigr)'(t).
\end{equation}
The weight $w$ is nonnegative (as $S$ is decreasing), satisfies $\int_0^\theta w(t)\,dt=1-S(\theta)/S(0)\le1$, and does not depend on $N$. Since the ratio $P_n(t)/P_n(\theta)$ lies in $[0,1]$, dominated convergence implies that $W_i^{m,N}(\theta)\to1$ if $P_n(t)/P_n(\theta)\to0$ for almost every $t<\theta$.

So it remains only to establish that the ratio does vanish, in fact at every $t<\theta$. The event that the types of at most $m-1$ of the $n$ opponents exceed $t$ has probability
\[
P_n(t)=\sum_{k=0}^{m-1}\binom{n}{k}\bigl(1-F(t)\bigr)^k F(t)^{n-k}\ \le\ m \cdot n^{m-1}F(t)^{n-m+1},
\]
while $P_n(\theta)\ge F(\theta)^n$, the probability that every opponent falls below $\theta$. Hence
\[
\frac{P_n(t)}{P_n(\theta)}\ \le\ \frac{m \cdot n^{m-1}}{F(\theta)^{m-1}}\left(\frac{F(t)}{F(\theta)}\right)^{n-m+1}\ \to \ 0 \text{ as } n\to \infty,
\]
because $F(t)<F(\theta)$ so that $\left(F(t)/F(\theta)\right)^{n-m+1}$ vanishes geometrically in $n$ while $n^{m-1}$ increases only polynomially.

\begin{remark}\label{rem:shutout}
The derivation of \eqref{e:rent-share} applies verbatim to any agent, with her own strain function $S_i$ and winning probability in place of $S$ and $P_n$. It thus yields a characterization of full dissipation well beyond symmetric agents. Concretely, hold agent $i$'s cost function fixed, consider any sequence of tournaments in which the opponents' type distributions and the number of prizes $m_N<N$ may change with $N$, and write $P^{N}_i$ for agent $i$'s winning probability. Then, for any $\theta>0$, the waste $W_i^{m_N,N}(\theta)\to1$ if and only if every type $t<\theta$ is eventually shut out of contention relative to type $\theta$, i.e., $P^{N}_i(t)/P^{N}_i(\theta)\to0$. The ``if'' direction is the dominated convergence argument above. For the ``only if,'' since $P^{N}_i(t)$ is increasing in $t$, \eqref{e:rent-share} implies
\[
1-W_i^{m_N,N}(\theta)\;\ge\;\frac{P^{N}_i(t)}{P^{N}_i(\theta)}\int_t^\theta w(u)\,du\;=\;\frac{P^{N}_i(t)}{P^{N}_i(\theta)}\left(1-\frac{S_i(\theta)}{S_i(t)}\right),
\]
whose second factor is positive (as $S_i$ is strictly decreasing) and independent of $N$; hence $W_i^{m_N,N}(\theta)\to1$ forces $P^{N}_i(t)/P^{N}_i(\theta)\to0$. Notably, whether full dissipation occurs does not depend on any agent's cost function.
\end{remark}

\subsection{Aggregate Dissipation}
\label{sec:aggregate-dissipation}

We next establish the aggregate dissipation statement referenced in the text after \autoref{prop:tournament}.

\begin{proposition}
\label{prop:aggregate-dissipation}
Consider the signaling tournament with $m$ prizes and $N$ symmetric agents, and let $A^{m,N}$ denote the common separating-equilibrium strategy. Holding $m$ fixed as $N\to\infty$, expected aggregate signaling costs converge to the full value of the prizes:
\[
\lim_{N\to \infty} \; N\cdot \E\bigl[C(A^{m,N}(\theta),\theta)\bigr] \;=\; s\cdot m.
\]
\end{proposition}

Because all $m$ prizes are always awarded, $N\cdot\E[V(\theta)]=s \cdot m$; the proposition thus says that, in expectation, aggregate signaling asymptotically dissipates the entire surplus.

\begin{proof}[Proof of \autoref{prop:aggregate-dissipation}]
Since all $m$ prizes are always awarded, $N \cdot \E[P_n(\theta)]=m$, where, as in the proof of \autoref{prop:tournament}, $P_n(\theta)$ is the probability that at most $m-1$ of the other $n=N-1$ types exceed $\theta$. So $d\mu_N(\theta):=(N/m)P_n(\theta)f(\theta)\,d\theta$ defines a probability measure on $[0,1]$, namely the distribution of the type of a uniformly selected prize winner. Since equilibrium costs are $C(A^{m,N}(\theta),\theta)=V(\theta)\,W_i^{m,N}(\theta)$ with $V=sP_n$, we have
\[
\frac{N \cdot \E\bigl[C(A^{m,N}(\theta),\theta)\bigr]}{s\cdot m}=\int_0^1 W_i^{m,N}(\theta)\,d\mu_N(\theta)\;\le\;1.
\]
Two observations complete the proof. First, $\mu_N$ concentrates at the top: for any $t<1$, a winner with type at most $t$ requires that fewer than $m$ of the $N$ types exceed $t$, which has vanishing probability as $N\to\infty$; hence $\mu_N$ converges weakly to the point mass at $\theta=1$. Second, waste is monotone in the field size: fixing any $N_0$, \autoref{prop:tournament} implies $W_i^{m,N}(\theta)\ge W_i^{m,N_0}(\theta)$ for all $N\ge N_0$ and $\theta>0$. Since $W_i^{m,N_0}$ is bounded and continuous at $\theta=1$, weak convergence yields
\[
\liminf_{N\to\infty}\int_0^1 W_i^{m,N}\,d\mu_N\;\ge\;\lim_{N\to\infty}\int_0^1 W_i^{m,N_0}\,d\mu_N\;=\;W_i^{m,N_0}(1).
\]
Now letting $N_0\to\infty$, \autoref{prop:tournament} also gives $W_i^{m,N_0}(1)\to1$. Hence $\int_0^1 W_i^{m,N}\,d\mu_N\to1$, i.e., $N \cdot \E\bigl[C(A^{m,N}(\theta),\theta)\bigr]\to s\cdot m$.
\end{proof}
\section{Waste Need Not be Constant in Stakes}
\label{sec:wasteandstakes}

The following result shows how the property that waste is constant in type can generalize beyond the isoelastic environment (part \ref{constant} of \autoref{thm:constant}), once we move beyond multiplicative costs. Unlike in the isoelastic environment, this constant waste can vary with stakes.

\begin{proposition}
\label{prop:constant-waste-moregeneral}
Assume $V(\theta)=sB(\theta)$ and 
\begin{align}
C(a,\theta)=B(\theta) \cdot h\left(\frac{D(a)}{Q(\theta)}\right), \label{eq:costforconstant}
\end{align} where $Q(0)=D(0)=h(0)=0$ with $Q',D',h'>0$.\footnote{Our \autoref{ass:basics} also imposes other properties, including the single-crossing condition, which under the constant relative elasticity assumption introduced momentarily reduces to $x h''(x)/h'(x)>\lambda-1$ for all $x>0$.} Denote the waste ratio at type $\theta$ under stakes $s$ by $W^s(\theta)$. If 
the relative elasticity $\frac{d\ln B(\theta)}{d\ln Q(\theta)}$ is constant at some $\lambda>0$, 
and if there is a unique solution $x^*(s)>0$ for $x$ in the equation $x h'(x)=s \lambda$, then waste is $W^s(\theta) = h(x^*(s))/s$, which is constant over $\theta>0$.
\end{proposition}

Under the proposition's conditions, using $s=x^*(s) h'(x^*(s)) / \lambda$, the constant waste can be written as 
\[
W^s(\theta)=\frac{\lambda h(x^*(s))}{x^*(s) h'(x^*(s))}=\frac{\lambda}{\varepsilon_h(x^*(s))},
\qquad
\text{where }\varepsilon_h(x):= \frac{d \ln h(x)}{ d \ln x} = \frac{x h'(x)}{h(x)}.
\]
Hence, since $x^*(s)$ is increasing in $s$ (differentiating and using the single-crossing condition, $(xh'(x))'>\lambda h'(x)>0$), waste is decreasing, increasing, or constant in stakes according to whether the elasticity $\varepsilon_h$ is increasing, decreasing, or constant.

In the \hyperref[def:isoelastic]{isoelastic environment}, $h$ is the identity and $\varepsilon_h(x)\equiv 1$, yielding  waste that is constant in both type and stakes.\footnote{Specifically, $B(\theta) = \theta^{\beta}$ and $C(a,\theta) = D(a) \theta^{-\sigma} = B(\theta) h\left(D(a) / Q(\theta)\right)$ for $h(x) = x$ and $Q(\theta) = \theta^{\sigma+ \beta}$.} Examples \ref{eg:nonmultsepcost}  and \ref{eg:countereg} below illustrate cases in which waste is constant across types, but is increasing or decreasing in the stakes. Example \ref{eg:nonmultsepcost}  also illustrates the result of \autoref{prop:comp_stat_nonmult} that an increase in cost elasticities will decrease the waste ratio.

\begin{proof}[Proof of \autoref{prop:constant-waste-moregeneral}]
Consider the strategy $A$ defined by $A(\typemin)=0$ and for $\theta>0$, $A(\theta)=D^{-1}\left(x^*(s) Q(\theta)\right)$. A routine computation shows that\footnote{Suppressing the argument of $x^*$, from $D(A(\theta))=x^*Q(\theta)$ we have $D'(A(\theta))A'(\theta)=x^*Q'(\theta)$, 
and hence
\[
C_a(A(\theta),\theta)A'(\theta)
=
B(\theta)h'(x^*)\frac{x^*Q'(\theta)}{Q(\theta)}=\frac{B'(\theta)}{\lambda}h'(x^*)x^*=sB'(\theta),
\]
where the second equality is because the constant relative elasticity assumption is equivalent to $Q'/Q=(1/\lambda)\cdot B'/B$, and the last equality is because $x^*h'(x^*)=s\lambda$.}
\[
C_a(A(\theta),\theta)A'(\theta)=sB'(\theta)=V'(\theta).
\]
Hence \autoref{e:foc} holds, and so $A$ is the separating equilibrium strategy.

Along the equilibrium path,
\[
C(A(\theta),\theta)
=
B(\theta) h\!\left(\frac{D(A(\theta))}{Q(\theta)}\right)
=
B(\theta) h(x^*(s)).
\]
Therefore, for $\theta>0$, the waste ratio is
\[
W^s(\theta)=\frac{C(A(\theta),\theta)}{V(\theta)}
=\frac{B(\theta) h(x^*(s))}{sB(\theta)}
=\frac{h(x^*(s))}{s}. \qedhere
\]
\end{proof}

\begin{example}
\label{eg:nonmultsepcost}
Let the benefit be linear with $V(\theta) = s \theta$ for $s>0$, and let the costs take the non-multiplicative form $C(a, \theta) = a^2 / (\tau \theta + a)$ for $\tau>0$, which satisfies \autoref{ass:basics}.

For small $a$, the cost is approximately $a^2/(\tau\theta)$, which is isoelastic with strain elasticity $1$. For large $a$, the cost is approximately $a$, which is independent of type. Higher stakes push agents to higher actions, for which high types have a smaller relative cost advantage. 
This suggests that higher stakes should increase waste, which we confirm below.

It is straightforward to verify using \autoref{e:foc} that the separating equilibrium has the linear strategy
$A^{\tau,s}(\theta)=\zeta(\tau,s)\tau \theta$, 
where $\zeta(\tau,s)>0$ is the unique value of $\zeta$ that solves
\begin{equation}
\label{e:nonmult-eg-compstats}
\frac{\zeta^2(2+\zeta)}{(1+\zeta)^2}=\frac{s}{\tau}.
\end{equation}
The constant waste is thus
\begin{equation} \label{eq:Wmsinexample}
W^{\tau,s}(\theta)=\frac{(\zeta(\tau,s)\tau\theta)^2}{\tau\theta+\zeta(\tau,s)\tau\theta}\cdot \frac{1}{s\theta}
=\frac{(\zeta(\tau,s))^2\tau}{s(1+\zeta(\tau,s))}
=\frac{1+\zeta(\tau,s)}{2+\zeta(\tau,s)}.
\end{equation}
As the left-hand side of \eqref{e:nonmult-eg-compstats} is strictly increasing in $\zeta$, we see that $\zeta(\tau,s)$ is strictly decreasing in $\tau$ and strictly increasing in $s$. Hence $W^{\tau,s}$ is strictly decreasing in $\tau$ and strictly increasing in $s$, with range $(1/2,1)$. 

Waste monotonicity in the parameter $\tau$ is an instance of \autoref{prop:comp_stat_nonmult} because $\elaB(\theta)=1$ is independent of $\tau$ while the on-path cost elasticity is
\[
\elaS^{A^{\tau,s}}(\theta)
=
-\theta \frac{C_\theta(A^{\tau,s}(\theta),\theta)}{C(A^{\tau,s}(\theta),\theta)}
=
\frac{1}{1+\zeta(\tau,s)},
\]
which is strictly increasing in $\tau$.

Connecting the result to \autoref{prop:constant-waste-moregeneral},
the cost $C(a,\theta) = a^2 / (\tau \theta + a)$ can be put in the form of \eqref{eq:costforconstant}  with $$h(x)=\frac{x^2}{\tau+x}, \quad B(\theta) = Q(\theta)=\theta, \quad D(a)=a.%
$$
The function $h$ is increasing, and there is a constant relative elasticity $d \ln B(\theta) / d \ln Q(\theta) = 1$. 
The proposition thus implies that waste equals $h(x^*(\tau,s))/s$ for $x^*(\tau,s)$ the solution to $x h'(x) = s$, i.e., $x^*(\tau,s)$ solving
\[\frac{x^2 (2\tau+x) }{ (\tau+x)^2}= s.\]
The relevant solution is $x^*(\tau,s)=\zeta(\tau,s)\tau$, yielding waste of 
\[\frac{h(x^*(\tau,s))}{s} = \frac{(\zeta(\tau,s))^2 \tau^2}{s (\tau + \zeta(\tau,s) \tau)} =\frac{(\zeta(\tau,s))^2 \tau}{s (1 + \zeta(\tau,s) )}, \]
just as in \eqref{eq:Wmsinexample}.
\end{example}

\begin{remark}Notice that in the separating equilibrium of \autoref{eg:nonmultsepcost}, the ratio $\elaB(\theta)/\elaS^A(\theta)$ is constant in type. Indeed, there is a generalization of \autoref{thm:char} to non-multiplicative costs: waste is constant in type if and only if the ratio $\elaB(\theta)/\elaS^A(\theta)$ is constant on $(\typemin,\typemax)$. The logic is in the proof of \autoref{prop:comp_stat_nonmult}: using \autoref{e:general-Wprime}, we see that $W'(\theta)=0$ implies 
$\elaB(\theta)/\elaS^A(\theta)=W(\theta)/(1-W(\theta))$; and so a constant waste implies a constant $\elaB(\theta)/\elaS^A(\theta)$. Conversely, if $\elaB/\elaS^{A}\equiv\rho$, then \eqref{e:general-Wprime} becomes $W'(\theta)=\frac{\elaS^{A}(\theta)}{\theta}\bigl(\rho-(1+\rho)W(\theta)\bigr)$, which admits the constant solution $\rho/(1+\rho)$; the integrating-factor argument in that proof, with its vanishing boundary term, shows the waste ratio must coincide with this constant solution.
\end{remark}

\begin{example}
\label{eg:countereg}
Let the benefit be linear with $V(\theta) = s \theta$ for $s>0$, and let the costs take the non-multiplicative form $C(a,\theta) = a^2/\theta + a^3/\theta^2$, which satisfies \autoref{ass:basics}.

For small $a$, the cost is approximately $a^2/\theta$, which is isoelastic with strain elasticity $1$. For large $a$, the cost is approximately $a^3 / \theta^2$, which is isoelastic with strain elasticity $2$. Higher stakes push agents to higher actions, for which high types have a larger relative cost advantage due to the higher strain elasticity. 
This suggests that higher stakes should decrease waste, which we confirm below.

It is straightforward to verify using \autoref{e:foc} that the separating equilibrium has the linear strategy $A^s(\theta) = c(s) \theta$, where $c(s) > 0$ is the unique solution to 
\begin{equation}
\label{e:eg-coefficient}
2c^2 + 3c^3 = s.
\end{equation}  

The waste ratio is thus
\begin{equation} \label{eq:increasingconstwaste}
W^s(\theta) = \frac{C(A^s(\theta),\theta)}{V(\theta)} = \frac{(c(s))^2 \theta + (c(s))^3 \theta}{\left[2(c(s))^2 + 3(c(s))^3\right]\theta} = \frac{1 + c(s)}{2 + 3c(s)},
\end{equation}
which is constant across types but depends on $s$. In particular, since $c$ is increasing in $s$ (\autoref{e:eg-coefficient}) and $W^s$ is decreasing in $c$, we see that $W^s$ is decreasing in $s$ with range 
$(1/3,1/2)$.

Connecting the result to \autoref{prop:constant-waste-moregeneral},
the cost $C(a,\theta) = a^2 / \theta + a^3 / \theta^2$ can be put in the form of \eqref{eq:costforconstant}  with  $$h(x)=x^2+x^3, \quad B(\theta) = Q(\theta)=\theta, \quad  D(a)=a. $$
The function $h$ is increasing, and there is a constant relative elasticity $d \ln B(\theta) / d \ln Q(\theta) = 1$, as in \autoref{eg:nonmultsepcost}.\footnote{This example's cost function and the one from \autoref{eg:nonmultsepcost} are special cases of $h(x) = x^2 (\tau+x)^l$, with $l=-1$ in \autoref{eg:nonmultsepcost} and with $\tau=1$ and $l=1$ here. A value of $l=0$ would yield the isoelastic cost $C(a,\theta) = a^2/\theta$.} The proposition thus implies that waste is equal to $h(x^*(s))/s$ for $x^*(s)$ the solution to $x h'(x) = s$, i.e., $x^*(s)$ solving $2x^2 + 3x^3 = s$. We see that $x^*(s)$ is identical to $c(s)$ from \eqref{e:eg-coefficient}, yielding waste of 
\[\frac{h(x^*(s))}{s} = \frac{c(s)^2 + c(s)^3}{s} =\frac{c(s)^2 + c(s)^3}{2c(s)^2 + 3 c(s)^3} = \frac{1+ c(s)}{2+3c(s)} \]
just as in \eqref{eq:increasingconstwaste}.
\end{example}

\section{Signaling Costs as Payments}
\label{sec:all-pay}

This appendix provides an alternative interpretation of some of our results by fleshing out the connection between separating equilibria and mechanism design. \appendixref{sec:payment-identity} details how equilibrium signaling costs can be viewed as payments. \appendixref{sec:allpay-rep} then shows that the signaling game is equivalent to an all-pay auction, and that translation affords an order-statistics reading of the waste ratio.

\subsection{The Payment Identity}
\label{sec:payment-identity}

Assume multiplicative costs and normalize the stakes to $s=1$, so that $V(\cdot)=B(\cdot)$. In a separating equilibrium, type $\theta$'s choice of an action can be equivalently viewed as choosing which type $\hat\theta$ to mimic, i.e., choosing $\hat \theta$ to maximize $B(\hat\theta)-D(A(\hat\theta))S(\theta)$. Divide this payoff by $S(\theta)>0$, which leaves incentives unchanged.\footnote{$S(\theta)$ is finite for all $\theta>\typemin$; if $S(\typemin)=\infty$, then assigning $v(\typemin)=1/\infty=0$ also preserves type $\typemin$'s incentives.} Writing
\[
v(\theta):=\frac{1}{S(\theta)}
\qquad\text{and}\qquad
b(\hat\theta):=D(A(\hat\theta)),
\]
type $\theta$'s problem becomes
\begin{equation}
\label{e:rescaled}
\max_{\hat\theta}\;\; v(\theta)\,B(\hat\theta)-b(\hat\theta),
\end{equation}
which is the familiar form in mechanism design where an agent with \emph{value} $v(\theta)$ chooses an \emph{allocation} $B(\hat\theta)$ with \emph{payment} $b(\hat\theta)$. Note that there is a bijection between types and values because $S$ is strictly decreasing.

In this notation, dividing the second equality of \eqref{e:eqm-cost} by $S(\theta)$ gives
\begin{equation}
\label{e:paymentidentity}
b(\theta)=\int_\typemin^\theta \frac{B'(t)}{S(t)}\,dt
=\int_\typemin^\theta v(t)\,dB(t)
=\underbrace{v(\theta)B(\theta)}_{\text{gross surplus}}-\underbrace{\int_\typemin^\theta B(t)\,dv(t)}_{\text{information rent}},
\end{equation}
where the last equality is from integration by parts, using $v(\typemin)B(\typemin)=0$. This is the \emph{payment identity} of mechanism design \citep{Myerson81}: a type's payment equals the gross surplus from her allocation less her information rent.

In the above transformation, one must bear in mind that $S(\theta)$ is a type-specific scaling. Thus the payment is not a type's signaling cost; instead, $b(\theta)=C(A(\theta),\theta)/S(\theta)$. Similarly for the gross surplus $v(\theta)B(\theta)$. However, the ratio has a clean interpretation: for $\theta>0$,
\begin{equation}
\label{e:wasteispayment}
W(\theta)=\frac{C(A(\theta),\theta)}{V(\theta)}
=\frac{b(\theta)}{v(\theta)B(\theta)}.
\end{equation}
In other words, the waste ratio is the payment's share of gross surplus.

\begin{remark}
\label{rem:relabel}
The discussion after \autoref{thm:char} shows that waste is constant exactly when $S=\kappa B^{-1/\rho}$. In the present language, this says the \emph{reward schedule} $x:=B\circ v^{-1}$ (the reward for an agent with value $v$) is isoelastic: $x(v)=c \cdot v^{\rho}$, with $W=\rho/(1+\rho)$. The ratio of elasticities $\rho$ in the theorem is thus the reward schedule's elasticity. An analogous change of variables, to quantiles rather than values, underlies the treatment of screening problems in \citet*{BergemannHeumannMorris26}. Indeed, their leading example is, in our notation, an isoelastic environment with $\beta=\sigma=4$.
\end{remark}

\subsection{An All-Pay Auction Representation}
\label{sec:allpay-rep}

The transformed payoff \eqref{e:rescaled} also delivers an all-pay auction representation of our single-agent signaling game. For simplicity assume $B$ is bounded (as is assured by $\typemax<\infty$), and normalize $\sup_{\hat \theta} B(\hat \theta)=1$ (a rescaling of stakes, which by \autoref{thm:constant} does not affect waste) so that $B(\hat \theta)$ can serve as a probability. Let $x=B\circ v^{-1}$ be the reward schedule, as in \autoref{rem:relabel}; $x$ is strictly increasing with $x(v(\typemin))=0$ and $\sup_{v} x(v)=1$. Fix any $N\ge2$ and define
\begin{equation}
\label{e:Gconstruct}
G(\hat v):=x(\hat v)^{1/(N-1)},
\end{equation}
a cumulative distribution function on the range of $v$. 
Consider a symmetric $N$-bidder independent-private-value all-pay auction in which values are distributed according to $G$, and suppose all bidders use a common differentiable, strictly increasing bidding strategy $\tilde b$ with $\tilde b(v(\typemin))=0$. A bidder with value $v$ who bids to mimic $\hat v$ wins with probability $G(\hat v)^{N-1}$ and pays $\tilde b(\hat v)$, so her payoff is
\begin{equation}
\label{e:auctionpayoff}
v\,G(\hat v)^{N-1}-\tilde b(\hat v).
\end{equation}
By construction, $G(\hat v)^{N-1}=x(\hat v)=B(\hat\theta)$ for $\hat v=v(\hat\theta)$, so when $\tilde b=b\circ v^{-1}$, the payoff \eqref{e:auctionpayoff} coincides with \eqref{e:rescaled}: the bidder's problem and the single agent's problem are identical. The pertinent equilibrium of the all-pay auction thus matches the signaling equilibrium; in particular, the equilibrium bid at value $v(\theta)$ is type $\theta$'s payment $b(\theta)=D(A(\theta))$, i.e., the separating strategy measured in difficulty units and indexed by value rather than type.

\begin{remark}
\label{rem:construction}
The equivalence above is between a single-agent signaling game and an all-pay auction construction in which the number of bidders $N$ is a free parameter and the original type distribution $F$ plays no role (as it does not affect the separating equilibrium). By contrast, the signaling tournament of \autoref{sec:tournament} has a given number $N$ of agents actually competing. With symmetric agents in the tournament, the equivalence extends to any number of prizes $m$: rescaling by $S(\theta)$ makes each agent a bidder with value $v(\theta)$ distributed according to $F\circ v^{-1}$, and awarding prizes to the $m$ highest inferred types is exactly the all-pay rule awarding $m$ identical objects to the $m$ highest bids. The benefit function then depends explicitly on the given $N$ and $F$, because an agent perceived as $\hat\theta$ wins a prize with the probability that at most $m-1$ of the other $N-1$ agents have types above $\hat \theta$.
\end{remark}

\paragraph{Waste and the runner-up.} The all-pay auction also provides another interpretation of the waste ratio.

\begin{proposition}
\label{prop:runnerup}
The waste ratio satisfies, for any number of bidders $N\ge2$ and type $\theta>0$,
\[
W(\theta)=\frac{\E\left[Y \;\middle|\; Y<v(\theta)\right]}{v(\theta)},
\]
where $Y$ is the (random) highest value among the other $N-1$ bidders.
\end{proposition}

\begin{proof}
By revenue equivalence \citep{Myerson81}, a bidder with value $v(\theta)$ bids her expected payment in the corresponding second-price auction: the probability $G(v(\theta))^{N-1}$ of winning times the expected runner-up value $\E\left[Y \mid Y<v(\theta)\right]$. This bid is type $\theta$'s payment $b(\theta)$; since $B(\theta)=G(v(\theta))^{N-1}$, the winning probability cancels from the payment's share of gross surplus \eqref{e:wasteispayment}, yielding the result.
\end{proof}

In words, the proposition says that an agent's waste is the expected runner-up's value as a fraction of the agent's own. While there is a genuine runner-up in a symmetric one-prize signaling tournament (\autoref{sec:tournament-simpler}), in the single-agent case it is fictitious, with the proposition then an accounting identity. Either way, waste is constant across types exactly when the expected runner-up is a constant fraction of the winner's value. Since the runner-up's value $Y$ 
satisfies $\Prob(Y \le v)=G(v)^{N-1}=x(v)$, 
isoelasticity ($x(v)=c \cdot v^{\rho}$) gives $\E\left[Y \mid Y<v\right]=v\cdot\rho/(1+\rho)$, recovering the constant waste formula \eqref{e:constant}, with $\rho=\beta/\sigma$ as in \autoref{thm:char}.

\bibliographystyle{aer}
\bibliography{FK-dissipation}

\newpage

\begin{center}
\textbf{{\LARGE \textcolor{DarkRed}{Supplementary Appendix}}}
\end{center}

\section{Omitted Proofs}
\label{sec:proof:separating-differentiable}

The following lemma is used in the proof of \autoref{prop:separating-differentiable}.

\begin{lemma}\label{lem:cost-type-zero}
$C(\cdot, 0)$ is strictly increasing on $\{a \geq 0 : C(a, 0) < \infty\}$.
\end{lemma}

\begin{proof}
Since $C(\cdot, \theta)$ is strictly increasing for $\theta > 0$, the limit definition of $C(\cdot, 0)$ implies it is weakly increasing. To establish strict monotonicity, consider $a'' > a' \geq 0$ with $C(a'', 0) < \infty$. For any $\theta > 0$, we have
\[
C(a'', \theta) - C(a', \theta) = \int_{a'}^{a''} C_a(x, \theta) \, dx.
\]
For any $x > 0$, the function $C_a(x, \cdot)$ is strictly decreasing on $(0,\typemax\rangle$ because $C_{a \theta}<0$ on this domain, and so $L(x) := \lim_{\theta \rightarrow 0} C_a(x, \theta)$ exists in $\Rstrpos \cup \{\infty\}$. By monotone convergence,
\[
C(a'', 0) - C(a', 0) = \int_{a'}^{a''} L(x) \, dx > 0. \qedhere
\]
\end{proof}

\begin{proof}[Proof of \autoref{prop:separating-differentiable}]
We proceed in three steps.

\medskip
\noindent\textsc{Step 1: Any separating equilibrium has a pure strategy.}

Consider any separating equilibrium strategy $\alpha$ and type $\theta$. Incentive compatibility requires  $C(a, \theta) = C(a', \theta)$ for $\alpha(\theta)$-a.e.~$a$ and $a'$, since any such actions induce the same belief $\theta$. By the strict monotonicity of $C(\cdot,\theta)$ for $\theta>0$ and using \autoref{lem:cost-type-zero} for $\theta=0$ (noting that this type will never choose an action with infinite cost), we have $a = a'$ for $\alpha(\theta)$-almost every $a$ and $a'$, which implies $\alpha(\theta)$ has singleton support. Hence $\alpha$ is a pure strategy.

\medskip
\noindent\textsc{Step 2: The boundary condition $A(0) = 0$.}

Let $A$ be a separating equilibrium (pure) strategy. If $A(\typemin)>0$, then $V(\typemin)-C(A(\typemin),\typemin)<V(\typemin)-C(0,\typemin)$, contradicting incentive compatibility (IC) for type $\typemin$.

\medskip
\noindent\textsc{Step 3: Monotonicity, continuity, and differentiability.}

Let $A$ be a separating equilibrium (pure) strategy. We first show $A$ is strictly increasing. Consider any $\theta' > \theta$. IC for type $\theta$ to not mimic $\theta'$ gives $C(A(\theta'), \theta) - C(A(\theta), \theta) \geq V(\theta') - V(\theta) > 0$. Since $C(\cdot, \theta)$ is nondecreasing, it follows that $A(\theta') > A(\theta)$.

We next show $A$ is continuous. Consider any $\theta^*\in (0,\typemax)$ and let $a^- := \lim_{\theta \uparrow \theta^*} A(\theta)$ and $a^+ := \lim_{\theta \downarrow \theta^*} A(\theta)$, which exist by monotonicity. IC for type $\theta^*$ implies that for all $\theta<\theta^*$, we have
\[
V(\theta^*) - C(A(\theta^*), \theta^*) \geq V(\theta) - C(A(\theta), \theta^*).
\]
Taking $\theta \uparrow \theta^*$ and using continuity of $V$ and $C$ yields
\begin{equation}
\label{e:continuity-IC1}
V(\theta^*) - C(A(\theta^*), \theta^*) \geq V(\theta^*) - C(a^-, \theta^*).    
\end{equation}
Applying IC in the reverse direction (for a type $\theta<\theta^*$ to not mimic $\theta^*$) and taking the same limit yields the opposite inequality to \eqref{e:continuity-IC1}. Hence $C(A(\theta^*), \theta^*) = C(a^-, \theta^*)$. An analogous argument using types above $\theta^*$ gives $C(A(\theta^*), \theta^*) = C(a^+, \theta^*)$. Since $C(\cdot, \theta^*)$ is strictly increasing, this implies $a^- = A(\theta^*) = a^+$, so $A$ is continuous at $\theta^*$. 
The same argument, using only $a^-$, also establishes continuity at $\theta^*=\typemax$ when $\typemax<\infty$. At $\theta^*=\typemin$, observe that if $a^+:=\lim_{\theta \downarrow \typemin} A(\theta)>A(\typemin)=0$, then some type close to  $\typemin$ would profitably deviate to action $0$ because (i) $V(\theta)\to 0$ as $\theta \to \typemin$, whereas (ii) for $\theta>\typemin$, $C(A(\theta),\theta)\geq C(a^+,\theta)$ is bounded away from $0$ because $C(a^+,\cdot)$ is decreasing. Hence $\lim_{\theta \downarrow \typemin}A(\theta)=A(\typemin)$.

We now establish differentiability on the interior. Fix $\theta\in(\typemin,\typemax)$ and $\theta' \neq \theta$. Adding the IC inequalities for $\theta$ to not mimic $\theta'$ and vice-versa, and rearranging, yields
\begin{equation}
\label{e:differentiability}
C(A(\theta'), \theta) - C(A(\theta), \theta) \geq V(\theta') - V(\theta) \geq C(A(\theta'), \theta') - C(A(\theta), \theta').
\end{equation}
Since $C$ is differentiable in its first argument, the mean value theorem implies that the left-hand side of \eqref{e:differentiability} equals $C_a(\tilde{a}, \theta) \cdot (A(\theta') - A(\theta))$ for some $\tilde{a}$ between $A(\theta)$ and $A(\theta')$, and analogously for the right-hand side with $C_a(\tilde{a}', \theta')$ for some $\tilde{a}'$ between $A(\theta)$ and $A(\theta')$. Substituting into \eqref{e:differentiability} and dividing by $\theta' - \theta$ gives
\begin{equation}
\label{e:differentiability-2}
C_a(\tilde{a}, \theta) \left( \frac{A(\theta') - A(\theta)}{\theta' - \theta}\right) \geq \frac{V(\theta') - V(\theta)}{\theta' - \theta} \geq C_a(\tilde{a}', \theta') \left(\frac{A(\theta') - A(\theta)}{\theta' - \theta}\right),    
\end{equation}
where the inequalities are written for $\theta'>\theta$ and would flip if $\theta' < \theta$. Either way, take $\theta' \to \theta$: since $\tilde{a}, \tilde{a}' \to A(\theta)$ by continuity of $A$, and $C_a$ is continuous, both $C_a(\tilde{a}, \theta)$ and $C_a(\tilde{a}', \theta')$ converge to $C_a(A(\theta), \theta) > 0$. Since the middle term of \eqref{e:differentiability-2} converges to $V'(\theta)$ and the two outer terms share the common factor $\frac{A(\theta') - A(\theta)}{\theta' - \theta}$ with coefficients converging to the same positive limit, we conclude that $A'(\theta)$ exists with
\[
A'(\theta) = \frac{V'(\theta)}{C_a(A(\theta), \theta)}.
\]
Note that if $V'$ is continuous on $(\typemin,\typemax)$, then $A'$ is continuous on that domain because $C_a$ and $A$ are both continuous on that domain and $C_a(A(\theta), \theta) > 0$ for $\theta>0$ (noting that $A(\theta)>0$ for $\theta>\typemin$ by separation).

\end{proof}

\begin{proof}[Proof of \autoref{prop:sufficiency}]
    Let $A$ have the stated properties. Note that for interior $\theta$, if $A(\theta)=0$ then $C_a(0,\theta)>0$: the derivative must exist and be nonzero to satisfy \eqref{e:ODE}, as $V'(\theta)>0$, and the derivative cannot be negative because $C(\cdot,\theta)$ is continuous and $C_a(a,\theta)>0$ for $a>0$. Using \autoref{ass:basics}, it follows that for all interior $\theta$, no matter the value of $A(\theta)$, both $V'(\theta)>0$ and $C_a(A(\theta),\theta)>0$, and hence \eqref{e:ODE} implies $A'(\theta)>0$. By continuity of $A$, it is strictly increasing on $[\typemin,\typemax\rangle$.

    We now verify that $A$ defines a separating equilibrium. Since any off-path action is met with belief $\hat \theta=0$, it is strictly worse than action $0$. So any type $\theta$ can be viewed as only choosing which type $\tilde{\theta}$ to mimic, with payoff 
\[
\Pi(\theta, \tilde{\theta}) := V(\tilde{\theta}) - C(A(\tilde{\theta}), \theta).
\]
Consider incentive compatibility among types in $(\typemin,\typemax\rangle$. For any such $\theta$ and any interior $\tilde\theta$, we have $A(\tilde\theta)>0$ (as $A$ is strictly increasing with $A(\typemin)=0$), and so, substituting $V'(\tilde\theta)=C_a(A(\tilde\theta),\tilde\theta)\,A'(\tilde\theta)$ from \eqref{e:ODE},
\[
\Pi_2(\theta,\tilde \theta)=V'(\tilde\theta)-C_a(A(\tilde\theta),\theta)\,A'(\tilde\theta)=A'(\tilde\theta)\left[C_a(A(\tilde\theta),\tilde\theta)-C_a(A(\tilde\theta),\theta)\right].
\]
Since $A'(\tilde\theta)>0$ for interior $\tilde\theta$ and $C_{a\theta}<0$, this expression is positive when $\tilde\theta<\theta$ and negative when $\tilde\theta>\theta$. Hence $\Pi(\theta,\cdot)$ is increasing on $(\typemin,\theta)$ and decreasing on $(\theta,\typemax)$, so $\tilde\theta=\theta$ maximizes $\Pi(\theta,\cdot)$ on $(\typemin,\typemax)$; continuity of $\Pi(\theta,\cdot)$ extends the comparison to $\tilde\theta=\typemax$ when the type space includes $\typemax$. In particular, type $\typemax$ does not gain from mimicking any lower type, and no type gains from mimicking $\typemax$.

Finally, we address type $0$. Consider an arbitrary other type $\theta>0$. We must show $\Pi(\theta, \theta) \geq \Pi(\theta, 0)$ and $\Pi(0, 0)\geq \Pi(0,\theta)$. Taking each in turn: 
\begin{enumerate}
\item Incentive compatibility on $(\typemin,\typemax\rangle$ implies $\Pi(\theta, \theta) \geq \Pi(\theta, \tilde{\theta})$ for all $\tilde{\theta} > 0$, while continuity of $V$, $A$ and $C(\cdot,\theta)$ imply $\Pi(\theta, \tilde{\theta}) \to \Pi(\theta, 0)$ as $\tilde{\theta} \to 0$. Hence, $\Pi(\theta, \theta) \geq \Pi(\theta, 0)$.
\item For any $\tilde \theta>0$, the previous point implies  $\Pi(\tilde{\theta}, \tilde{\theta}) \geq \Pi(\tilde{\theta}, 0)=0$, and hence $0\leq C(A(\tilde{\theta}), \tilde{\theta}) \leq V(\tilde{\theta})$. Since  $V(\tilde \theta) \rightarrow 0$ as $\tilde{\theta} \rightarrow 0$, it follows that $\Pi(\tilde{\theta}, \tilde{\theta}) \rightarrow 0=\Pi(0,0)$. Moreover, for any $\tilde{\theta}>0$ we have $\Pi(\tilde{\theta}, \tilde{\theta}) \geq \Pi(\tilde{\theta}, \theta)$ by incentive compatibility on $(\typemin,\typemax\rangle$, and $\Pi(\tilde{\theta}, \theta) \rightarrow \Pi(0, \theta)$ by the limit property of $C(\cdot, 0)$ in \autoref{ass:basics}. Hence $\Pi(0,0) \geq \Pi(0,\theta)$.  \qedhere
\end{enumerate}
\end{proof}
\end{document}